\documentclass{siamonline190516}
\usepackage{amsmath}
\usepackage{amssymb}
\usepackage{tikz}
\usepackage{amsfonts}
\usepackage{algorithm}
\usepackage[noend]{algpseudocode}
\usepackage{adjustbox}

\usepackage{enumerate}
\usepackage{graphicx}
\usepackage{pifont}
\usepackage{tabularx}
\usepackage{multirow}
\usepackage[caption=false]{subfig}
\def\rsls {r\textsc{SLS}}
\usepackage{todonotes}
\usepackage{changebar}
\usepackage{easyReview}

\title{Estimating leverage scores via rank revealing methods and randomization}
\author{Aleksandros Sobczyk\thanks{\email{obc@zurich.ibm.com}, IBM Research Zurich, Sa\"{u}merstrasse 4, R\"{u}schlikon, Switzerland.}
\and
Efstratios Gallopoulos\thanks{\email{stratis@ceid.upatras.gr}, HPCLab, Computer Engineering \& Informatics Dept., University of Patras, 26504 Patras, Greece.}}
\headers{Leverage estimation via randomization and RRQR}{Aleksandros Sobczyk, Efstratios Gallopoulos}

\begin{document}

\maketitle

\begin{abstract}  

We study algorithms for estimating the statistical leverage scores of rectangular dense or sparse matrices of arbitrary rank. Our approach is based on combining rank revealing methods with compositions of dense and sparse randomized dimensionality reduction transforms. We first develop a set of fast novel algorithms for rank estimation, column subset selection and least squares preconditioning. We then describe the design and implementation of leverage score estimators based on these primitives. These estimators are also effective for rank deficient input, which is frequently the case in data analytics applications. We provide detailed complexity analyses for all algorithms as well as meaningful approximation bounds and comparisons with the state-of-the-art. We conduct extensive numerical experiments to evaluate our algorithms and to illustrate their properties and performance using synthetic and real world data sets. 
\end{abstract}


\section{Introduction\label{sec:introduction}}

In this work we develop algorithms for estimating the statistical leverage scores of large rectangular dense or sparse matrices of arbitrary rank. These are ubiquitous quantities in large scale graph analytics \cite{jindal2017density, newman2005measure, Spielman:2008ka}, data analytics  \cite{candes2007sparsity, chatterjee2015regression, talwalkar2010matrix, velleman1981efficient}, numerical linear algebra  \cite{Bhojanapalli:2015ea, boutsidis2009improved, Chen:2015:CAL:2789272.2912096, Cohen:2015cb, cohen2017input, drineas2018structural, drineas2012fast, durfee2017determinant, papailiopoulos2014provable} and machine learning  \cite{pmlr-v95-chin18a, gittens2013revisiting, perry2016augmented}. See \cite{musco2018spectrum} for a fine-grained complexity analysis and connections between leverage scores and related linear algebra and graph problems. The algorithms described are especially effective for matrices $A\in \mathbb{R}^{n \times d}$ with $n \gg d$, which are sometimes characterized by anthropomorphic terms as  ``tall-and-thin'' or  ``tall-and-skinny''.
We show that our algorithms have lower complexities than existing state-of-the-art methods, that they are fast in practice, and that the results obtained come with theoretical approximation guarantees that we validate experimentally. 
We outline our implementation on a shared memory parallel system, which is publicly available on Github \footnote{https://github.com/IBM/pylspack}, 
and illustrate its effectiveness on synthetic and real world datasets, most notably computing the leverage scores of a sparse matrix of size $80$ million$\times 1,024$ containing $\mathcal{O}(10^9)$  nonzeros.

We recall that the \textit{row statistical leverage scores} (\rsls\ for short) of a tall matrix are defined as the Euclidean lengths of the rows of the matrix of any orthogonal basis for ${\rm range}(A)$, that is the column subspace of $A$. A mathematically equivalent definition is that the leverage scores are the elements in the diagonal, ${\theta}={\rm diag}(H)$, of the ``hat matrix'' 
$H=A (A^\top A)^{\dagger} A^\top$
so that
\begin{equation}
    \theta_i = \|e_i^\top A\|_{(A^\top A)^{\dagger}}, \forall i\in[n]
    \label{eq:ls_orthonormal_basis}
\end{equation}
where $\|c\|_M=c^\top M c$. Note that the leverage scores can also be expressed as 
\begin{equation}
    \theta_i = \|e_i^\top AX\|^2, X=\arg\min_Y\|AY-I_n\|_F,\forall i\in[n]
    \label{eq:ls_least_squares}
\end{equation}

The hat matrix is the (unique) orthogonal projector on the  column subspace of $A$ and is therefore also equal to $U U^\top$ for any matrix $U$ whose columns are an orthogonal basis for ${\rm range}(A)$. 
In a linear regression model ${ y}=A{x}+{r}$, for example,  the  hat matrix values reveal 
sensitive points in the design; the diagonal ones, in particular, reveal the leverage of each observed (response)  value in ${ y}$ on the corresponding fitted value in  ${ \hat{y}} = H{ y}$; cf. \cite{hoaglin1978hat}.  
In randomized numerical linear algebra (RandNLA), 
statistical leverages are important indicators in selection strategies for various algorithms\footnote{Without loss of generality, our discussion in the sequel is directed towards the computation of row leverage scores of tall-and-thin matrices. 
All our results are trivially adaptable to  column leverage scores of ``short-and-wide'' matrices.}, \cite{boutsidis2009improved,cohen2017input,Li:2013is,papailiopoulos2014provable}. 
Another quantity of interest (see e.g. \cite{avron2010blendenpik}), ``matrix coherence'', is also readily available from the \rsls\ since it 
is defined as the maximum leverage score. 

Exploiting the  Sherman-Morrison formula and the fact that $I_n-{ e}_i { e}_i^\top$ is an orthogonal projector, where $I_n$ is the identity matrix of size $n\times n$ and ${ e}_i$ is its $i$-th column, it is easy to show that if the covariance matrix
corresponding to $A$ 
is denoted by $G=A^\top A$ and the covariance corresponding to $A$ after deleting row $i$ is denoted by $G_{(i)}$, then
\begin{equation} \label{eq:sm}
G_{(i)}^{-1} =
G^{-1} + \frac{1}{(1-\theta_i)} G^{-1}A_{i,:} A_{i,:}^\top G^{-1}
\end{equation}
where $\theta_i$ is the corresponding \rsls\ value \cite{hoaglin1978hat} and $A_{i,:}$ is the $i$-th row of $A$ as a column vector. 
As it is known that $\theta_i \in [\frac{d}{n},1]$ for all \rsls, it follows from (\ref{eq:sm}) that  ignoring rows with \rsls\ close to 1, can 
cause large disturbances to the new covariance matrix. 

Computing the \rsls\ or even the entire hat matrix can be accomplished with few commands of  a MATLAB-like environment. In Julia \cite{Bezanson:2017gd}, for example,

{\tt H = A*((A'*A)} $\backslash$ {\tt A');} ${\rm \theta}$ {\tt = diag(H)}

\noindent or from an orthogonal basis for  ${\rm range}(A)$ via QR, e.g. 

{\tt Q,R = qr(A);}${\rm \theta}$={\tt sum(Q.*Q, dims=2);}.

The numerical issues that arise in deploying the direct methods that are feasible for smaller problems are discussed in
\cite{holodnak2015conditioning,Ipsen:2014ta}. The problem becomes  computationally challenging and a performance bottleneck when the matrix is very large, which is the case in ``big data'' analytics for which the above commands and associated standard solvers become prohibitively expensive. Moreover, special treatment is required when the input is approximately rank deficient, which is a common case in real applications. 

\subsection{Notation}

Throughout the paper,  unless stated otherwise,  the following notation is used. We follow the Householder notation, denoting matrices with capital letters, vectors with small letters, and scalars with Greek letters. All vectors are assumed to be columns.
$I_n$ is the identity matrix of size $n\times n$, ${ e}_i$ is the $i$-th column of the standard basis  and $0_{m\times n}$ is a zero matrix of size $m\times n$. For a matrix $A\in\mathbb{R}^{n\times d}$, $A_{i,:}$ and $A_{:,j}$ are the $i$-th row  and $j$-th column, respectively, both assumed to be column vectors, and $A_{i,j}$ is the element in row $i$ and column $j$. For a set $\mathcal{K}\subseteq [d]$,  $A_{:,\mathcal{K}}$ denotes the submatrix of $A$ containing the columns defined by $\mathcal{K}$. We denote by $\tilde A$ a ``sketch'' of $A$, defined as $\tilde A=SA$ for some suitably chosen matrix $S$, such that $SA$ approximates various properties of $A$ but has much smaller size. $A_k$ denotes the best rank-$k$ approximation of $A$ in the $2$-norm. $A^\dagger$ denotes the pseudoinverse.  Unless specifically noted otherwise, the 2-norm is assumed for matrices and vectors.
$[n]$ is the set $\{1,2,...,n\}$, where $n\in \mathbb{N}$. $\mathbb{P}[\alpha]\in[0,1]$ is the probability of occurrence of an event $\alpha$, while $\cal N$ $(\mu,\sigma)$ is the normal distribution with mean value $\mu$ and standard deviation $\sigma$. As usual, $\sigma_i(A)$ denotes the $i$-th largest singular value of $A$. The matrix argument is omitted when it is evident from the context. ${\tt nnz}(x)$ is the number of nonzeros of $x$, where in this case $x$ can be either a vector or a matrix. We also  define  ${\tt nnz_2}(A)=\sum_{i=1}^{n}\left({\tt nnz}(A_{i,:})\right)^2$, which will be used in complexity bounds, and $\tilde O(k):=\mathcal{O}(k\text{ polylog}(k))$. We will be referring to matrices with i.i.d. elements from $\mathcal{N}(0,1)$ as Gaussian matrices.

\subsection{Contributions}

We develop tools and algorithms for computing leverage scores based on Equation \ref{eq:ls_orthonormal_basis}.

All the algorithms discussed can be grouped as a template in a generic framework consisting of three basic steps, which we list in Algorithm \ref{alg:rsls_framework}.
\begin{algorithm}
\caption{Framework for estimating the \rsls\ of tall-and-thin matrices}
\label{alg:rsls_framework}
\begin{algorithmic}[1]
    \State Construct a submatrix $A_{:,\mathcal{K}}$ by selecting a subset of columns of $A$ defined by $\mathcal{K}\subseteq[d]$.
    \State Compute an ``orthogonalizer'' $R$ such that $A_{:,\mathcal{K}}R^{\dagger}$ has orthonormal columns.
    \State Compute the squared row norms of $A_{:,\mathcal{K}}R^{\dagger}$.
\end{algorithmic}
\end{algorithm}
\\
A key observation is that any algorithm based on this framework cannot be faster than the fastest algorithm for computing the squared row norms of the product $A_{:,\mathcal{K}}R^\dagger$, which, as Lemma \ref{lem:row_norms_complexity} suggests, can be done in $\mathcal{O}({\tt nnz_2}(A_{:,\mathcal{K}}))$\footnote{Note that for a dense matrix $A\in \mathbb{R}^{n\times d}$ we have that ${\tt nnz_2}(A)=nd^2$.}. This sheds new light in the design of leverage scores algorithms, as it turns out that, for sparse matrices, we can even form the entire Gram matrix $A_{:,\mathcal{K}}^\top A_{:,\mathcal{K}}$ in Equation \ref{eq:ls_orthonormal_basis} at the same cost (Lemma \ref{lem:gram_complexity}). To overcome the complexity bottleneck of the row norms computation we consider the following question: Is it possible to use only $k<\text{rank}(A)$ columns of $A$ in order to obtain good leverage scores approximations? We show that (Theorem \ref{thm:ls_qr_vs_svd}) if $n\gg d$ and we keep only $k<\text{rank}(A)$ columns  
then the following bound holds
\begin{align*}
|\theta_i(A_k) - \theta_i(A_{:,\mathcal{K}})|
\leq
\left(\sqrt{\theta_i(A_k)} + \sqrt{\theta_i(A_{:,\mathcal{K}})}
\right)
\dfrac{\sigma_{k+1}(A)}{\sigma_{k}(A_{:,\mathcal{K}})}.
\end{align*}
To keep the bound as small as possible, we need to select $A_{:,\mathcal{K}}$ such that $1/\sigma_k(A_{:,\mathcal{K}})=\|A_{:,\mathcal{K}}^\dagger\|$ is minimized, which is a variant of the Column Subset Selection Problem (CSSP) \cite{avron2017faster,boutsidis2009improved,civril2014column,shitov2021column}.

\subsubsection*{Rank estimation and column subset selection} 

We present a novel approach (Algorithm \ref{alg:count_gauss_strong_rrqr}) 
for column selection for tall-and-thin matrices.
Specifically, given a tall-and-thin matrix $A\in\mathbb{R}^{n\times d}$ 
we show a method for selecting a subset $\mathcal{K}\subseteq [d]$ with $|\mathcal{K}|=k<\text{rank}(A)\leq d$ in $\mathcal{O}({\tt nnz}(A)+d^4)$ operations, such that, with high probability it holds that $\sigma_k( A_{:,\mathcal{K}})>\frac{\sigma_k(A)}{C}$, for some bounded constant $C>1$ (Lemma \ref{lem:bound_sigma_min_r_11}).
This 
allows us to use this algorithm in the context of leverage scores to derive fast and provably good approximations (Theorem \ref{thm:ls_qr_vs_svd}). 
In Table \ref{tab:rank_estimators_comparison} we compare our approach with the algorithm of \cite{cheung2013fast}, which was utilized in leverage scores algorithms as a preprocessing step \cite{clarkson2013low_jacm,nelson2013osnap}, in order to identify the true rank of the matrix and select a set of rank$(A)$ columns of $A$. While the complexity of the two algorithms is similar, our approach has the advantage that it can also select $k<\text{rank}(A)$ columns to further reduce the input size.
In Section \ref{sec:sketching_cost_analysis} we show how the complexity can be reduced even further.
\begin{table}[htb]
    \centering
    \caption{Algorithms for selecting a submatrix $A_{:,\mathcal{K}}$ of a tall-and-thin matrix $A\in\mathbb{R}^{n\times d}$, where $\mathcal{K}\subseteq [d]$ with $|\mathcal{K}|=k\leq \text{rank}(A)\leq d$.  Here $m\in(d,3d]$, $\phi>1$ and $\epsilon,\delta\in(0,1)$ are input parameters. Also $\xi = \frac{(1+\alpha+\sqrt{k/m})}{(1-\alpha-\sqrt{k/m})}$, $\eta=\frac{(1+\epsilon)}{(1-\epsilon)}$, $\rho=\sqrt{1+\phi^2k(d-k)}$, and $\alpha\in(0,1-\sqrt{k/m})$.}
    \label{tab:rank_estimators_comparison}
    {\small
    \begin{tabular}{c  c  c  c}\hline
        Algorithm 
        & 
        Complexity 
        & 
        Success prob. 
        & 
        Approx. Bounds 
        \\\hline
        \cite[Thm. 2.11]{cheung2013fast}, Cor. \ref{cor:cheung_complexity_tall_and_thin}
        & 
        $\mathcal{O}({\tt nnz}(A)+d^\omega)$ 
        & 
        $1-\mathcal{O}(d^{-1/3})$
        & 
        - 
        \\
        Alg. \ref{alg:count_gauss_strong_rrqr}, Lem. \ref{lem:bound_sigma_min_r_11}
        &
        $\mathcal{O}({\tt nnz}(A)+md^3/(\epsilon^2\delta))$ 
        &
        $(1-\delta)(1-2e^{-\alpha^2m/2})$
        &
        $\sigma_{k}(A_{:,\mathcal{K}}) \geq \dfrac{\sigma_k(A)}{\xi\eta \rho}$
        \\\hline
    \end{tabular}
    }
\end{table}

\subsubsection*{Leverage scores}

Section \ref{sec:ls_algorithms} is devoted to the main objective of this paper, which is the computation of leverage scores of tall-and-thin matrices. Four novel algorithms are presented, their key properties being listed in Table \ref{tab:ls_estimators_comparison}. 
The first algorithm (Alg. \ref{alg:ls_direct}) directly computes Equation \ref{eq:ls_orthonormal_basis} and we use it as a baseline. Despite its simplicity, we show that it achieves the same complexity as state-of-the-art estimators \cite{clarkson2013low_jacm,nelson2013osnap} when the underlying matrix is not dense\footnote{These algorithms are extensions of \cite[Alg. 1]{drineas2012fast} for sparse matrices, which was originally designed for dense matrices.} (Lemmas \ref{lem:row_norms_complexity}, \ref{lem:gram_complexity}). To the best of our knowledge, this has not been observed before. 

We then show how we can improve existing algorithms and design new ones using our algorithm for column subset selection. In Theorem \ref{thm:ls_hrn_exact_bounds}, we show that we can select a submatrix of $A$ with $k<$rank$(A)$ columns, whose leverage scores will be close to the leverage scores of  $A_k$. 
We also describe a sparse pivoted QR-based approach \cite{stewart1999four}, which is especially  competitive for small ranks.
In summary, the algorithms described in this paper have the following properties. Given $A\in\mathbb{R}^{n\times d}$, $n\gg d$, they can
\begin{itemize}
    \item[--] compute the leverage scores of $A_k$ in $\mathcal{O}({\tt nnz_2}(A)+d^3)$ (Alg. \ref{alg:ls_direct}).
    \item[--] approximate the leverage scores of $A_k$ in $\mathcal{O}({\tt nnz}(A)+d^4+{\tt nnz_2}(A_{:,\mathcal{K}}))$ with high probability, where $\mathcal{K}\subseteq [d]$, $|\mathcal{K}|=k$ (Alg. \ref{alg:ls_hrn_exact}, Thm. \ref{thm:ls_hrn_exact_bounds}).
    \item[--] approximate the leverage scores of $A_k$ in $\mathcal{O}({\tt nnz}(A)+d^4+{\tt nnz}(A_{:,\mathcal{K}})\frac{\ln n}{\epsilon^2})$ with high probability, where $\epsilon\in(0,1)$ and $\mathcal{K}\subseteq [d]$, $|\mathcal{K}|=k$  (Alg. \ref{alg:ls_hrn_approx}, Thm. \ref{thm:ls_hrn_approx_bounds}).
\end{itemize}
In Section \ref{sec:ls_overall} the algorithms are ranked with respect to their complexity: Algorithms \ref{alg:ls_hrn_exact} and \ref{alg:ls_hrn_approx} appear to be the fastest for sparse and dense matrices respectively.

\begin{table}[htb]
    \centering
    \scriptsize
    \caption{Leverage scores algorithms for tall-and-thin matrices. 
    The set $\mathcal{K}\subseteq [d]$, $|\mathcal{K}|=k$ is selected by each algorithm individually.
    $\tilde \theta_i$ are the leverage scores returned by each algorithm
    and $\theta_i(A_k)$ are the leverage scores of $A_k$. Here $\delta,\epsilon\in(0,1)$ as well as $\psi\in(0,1]$ and  $m\in(d,3d]$ are input parameters, and $\alpha\in(0,1-\sqrt{k/m})$. Algorithms marked with $^\ddag$ return approximate solutions, and they are only faster than the exact algorithms when the matrix is sufficiently dense, specifically when ${\tt nnz_2}(A_{:,\mathcal{K}}) > {\tt nnz}(A_{:,\mathcal{K}})\frac{\ln n}{\epsilon^{2}}$ or ${\tt nnz_2}(A) > {\tt nnz}(A)\psi^{-1}$ accordingly.
    }
    \begin{tabular}{c  c  c  c c}
    \hline
    \\[-1em]
        \multirow{2}{*}{Algorithm }
        & 
        \multirow{2}{*}{Complexity }
        & 
        \multirow{2}{*}{Success prob. }
        & 
        \multicolumn{2}{c}{$\left|\tilde \theta_i - \theta_i(A_k)\right|$} 
        \\[+0.5em]\cline{4-5}
        & 
        & 
        & 
        $k=\text{rank}(A)$ & $k<$rank$(A)$ 
    \\\hline
    \\[-2em]
        & & & &
    \\
    
        Alg. \ref{alg:ls_direct} 
        & 
        $\mathcal{O}({\tt nnz_2}(A)+d^3)$
        & 
        deterministic
        & 
        exact
        &
        exact
    \\
        Alg. \ref{alg:ls_spqr} 
        & 
        $\mathcal{O}\left({{\tt nnz_2}(A_{:,\mathcal{K}})+}{\tt nnz}(A)(1+\frac{k^2}{d})\right)$
        & 
        deterministic
        & 
        exact
        &
        -
    \\
        Alg. \ref{alg:ls_hrn_exact} 
        & 
        $\mathcal{O}\left(
        {\tt nnz}(A)
        +
        md^3 
        +
        {\tt nnz_2}(A_{:,\mathcal{K}})
        \right)$
        & 
        $(1-\delta)(1-2e^{-\frac{\alpha^2m}{2}})$
        & 
        exact
        &
        Thm. \ref{thm:ls_hrn_exact_bounds}
    \\\hline
        Alg. \ref{alg:ls_hrn_approx}$^\ddag$
        &
        $\mathcal{O}\left(
        {\tt nnz}(A)+\frac{md^3}{\epsilon^2\delta}+ 
        {\tt nnz}(A_{:,\mathcal{K}})\frac{\ln n}{\epsilon^{2}}\right)$
        & 
        $(1-\delta)(1-2e^{-\frac{\alpha^2m}{2}})$
        &
        $\epsilon\theta_i (A_k)$
        &
        Thm. \ref{thm:ls_hrn_approx_bounds}
    \\
        \ref{alg:ls_approximate} \cite{clarkson2013low_jacm,nelson2013osnap}$^\ddag$
        & 
        $\mathcal{O}\left(
        {{\tt nnz}(A)}+d^\omega+ 
        {\tt nnz}(A_{:,\mathcal{K}})\frac{\ln n}{\epsilon^{2}}
        \right)$
        & 
        $(1-\delta)$
        & 
        $\epsilon\theta_i (A_k)$
        &
        -
    \\
        \cite[Alg. 2]{Cohen:2015cb}$^\ddag$
        & 
        $\mathcal{O}\left(
        \frac{{\tt nnz}(A)}{\psi}
        \right)
        +\tilde O\left(d^\omega+\frac{d^{2+\psi}}{\epsilon^2}\right)$
        & 
        high in $d$
        & 
        ${\mathcal{O}(d^\psi(1+\epsilon))}\theta_i(A_k)$
        &
        -
    \\\hline
    \end{tabular}
    \label{tab:ls_estimators_comparison}
\end{table}

\subsubsection*{Least squares preconditioning}
{In the course of establishing algorithms for leverage scores based on Equation \ref{eq:ls_orthonormal_basis}, we also obtained as byproduct effective randomized preconditioners for iterative least squares solvers.  While similar concepts have been discussed in \cite{clarkson2013low_jacm,woodruff2014sketching}, our work appears to be the first to provide a detailed  analysis and also systematically address the issue of parameter selection and experimental evaluation}.
Our results are summarized in Table \ref{tab:preconditioner_estimators_comparison}.

\begin{table}[htb]
    \centering
    \caption{
        Algorithms for obtaining least squares preconditioners for a rectangular matrix $A\in\mathbb{R}^{n\times d}$ with $n\gg d$. {The numerical rank $k\leq\text{rank}(A)\leq d$ is computed by each algorithm independently. Here $m\in(d,3d]$} and $\delta,\epsilon\in(0,1)$ are input parameters. Also $\alpha\in(0,1-\sqrt{k/m})$, $\xi = \frac{(1+\alpha+\sqrt{k/m})}{(1-\alpha-\sqrt{k/m})}$ and  $\eta=\frac{(1+\epsilon)}{(1-\epsilon)}$. $\tilde A$ denotes a ``sketch'' of $A$, used to construct the preconditioner $N\in\mathbb{R}^{d\times k}$. }
    \label{tab:preconditioner_estimators_comparison}
    {
    \footnotesize
    \begin{tabular}{c  c  c  c  c }\hline
        Algorithm 
        & 
        Complexity 
        & 
        Success prob. 
        &
        Bounds 
        &
        size of $\tilde A$
        \\\hline
        \cite[Alg. 1, steps 3-5]{meng2014lsrn}
        &
        $\mathcal{O}\left({\tt nnz}(A)m+md^2\right)$
        & 
        $(1-2e^{-\alpha^2m/2})$
        &
        $\kappa_2(AN)\leq \xi $ 
        &
        $m\times d$
        \\
        \cite[Alg. 2]{dahiya2018empirical}, \cite[\S 7.7]{clarkson2013low_jacm}
        &
        $\mathcal{O}\left({\tt nnz}(A)+\dfrac{d^4}{\epsilon^2\delta}\right)$
        & 
        $(1-\delta)$
        &
        $\kappa_2(AN)\leq \eta $
        &
        {$\Omega(\frac{d^2}{\epsilon^2})\times d$}, \cite{nelson2013sparsity}
        \\
        Alg. \ref{alg:count_gauss_precondition}, Thm. \ref{thm:cond_gsa}
         &
         $\mathcal{O}\left({\tt nnz}(A)+{\dfrac{md^3}{\epsilon^2\delta}}\right)$
         &
         $(1-\delta)(1-2e^{-\alpha^2m/2})$
         &
         $\kappa_2(AN)\leq \xi\eta$
         &
         $m\times d$
        \\\hline
    \end{tabular}
    }
\end{table}

\subsubsection*{Implementation and numerical experiments}
 In Section \ref{sec:experiments} we present experimental results, validating the theoretical approximation guarantees. We also show that the proposed methods can be implemented efficiently and perform well in practice on large datasets.

\section{Basic properties}
We start with a key observation regarding the computational complexity of leverage scores algorithms based on Algorithm \ref{alg:rsls_framework} and then proceed to discuss the case of leverage scores of rank deficient matrices.

\subsection{The ``curse'' of the row norms computation\label{sec:curse}}
 As already described above, an orthonormal base $U$ for range($A$) can be expressed as the product $U=AB$ where $B\in\mathbb{R}^{d\times d}$ is an ``orthogonalizer'' for $A$. Then, the leverage scores can be computed as \begin{align*}\theta_i=\|e_i^\top A\|_{(B B^\top)}^2=e_i^\top A (B B^\top) A^\top e_i\end{align*}
or approximated by 
\begin{align*}\theta_i=\|e_i^\top A\|_{(B\Pi \Pi^\top B^\top)}^2=e_i^\top A (B\Pi \Pi^\top B^\top) A^\top e_i\end{align*}
where $\Pi$ is a matrix with $r<d$ columns such that the row norms of $AB\Pi$ are approximately equal to those of $AB$. We state the following Lemma.
\begin{lemma}
\label{lem:row_norms_complexity}
Let $A\in\mathbb{R}^{n\times d}$, $n\gg d$ and $B\in\mathbb{R}^{d\times r}$, $r\leq d$. The squared Euclidean norms of the $n$ rows of the matrix $AB$ can be computed in 
$
\mathcal{O}
\left(
    \min\{
        {\tt nnz}(A)r,
        {\tt nnz_2}(A)+d^2r
    \}
\right)
$ operations.
\begin{proof}

Let {$A_{i,:}=A^\top e_i$} be the $i$-th row of $A$ as a column vector. The squared Euclidean norm of ${A_{i,:}^\top} B$ is equal to \begin{align*}
\|{A_{i,:}^\top} B\|^2 
= \left|
        \sum_{j=1}^{r}
        \left(\sum_{{\{k | A_{i,k}\neq 0\}}}
        {A_{i,k}}
        B_{k,j}
        \right)^2
    \right|
\end{align*}
or equivalently
\begin{align*}|{A_{i,:}^\top} (BB^\top) {A_{i,:}}| = \left|
        \sum_{{\{j | A_{i,j}\neq 0\}}}
        \sum_{{\{k | A_{i,k}\neq 0\}}}
        {A_{i,j} A_{i,k}} (BB^\top)_{k,j}
    \right|.
\end{align*}
The first sum requires $\mathcal{O}({\tt nnz}(A_{i,:})r)$ operations while the second sum requires $\mathcal{O}({\tt nnz}^2(A_{i,:}))$ operations, assuming that we have pre-computed $BB^\top$ in $\mathcal{O}(d^2r)$ operations. Therefore, the total number of operations required to compute all the sums over all the $n$ rows is
$\mathcal{O}
\left(
    \min\{
        {\tt nnz}(A)r,
        {\tt nnz_2}(A)+d^2r
    \}
\right)$.
\end{proof}
\end{lemma}

With this formulation, each $\theta_i$ can be computed in $\mathcal{O}({\tt nnz}^2(A_{i,:}))$, and all $\theta_i, i\in[n]$ in $\mathcal{O}({\tt nnz_2}(A))$ total arithmetic operations. Therefore, any algorithm that is based on constructing an orthogonalizer $B$ and computing the squared row norms of $AB$, cannot achieve a better complexity than $\mathcal{O}({\tt nnz_2}(A))$, unless either $A$ is sufficiently dense, or if $B$ is such that $\theta_i=\|e_i^\top A\|_{(B B^\top)}^2$ can be computed fast, with less than $\mathcal{O}({\tt nnz}^2(A_{i,:}))$ operations, e.g. if $B$ has either very few columns or if it has a special structure. In the context of existing leverage scores algorithms (both exact and approximate), none of these assumptions hold  for $B$, which typically has at least $\min\{d,\mathcal{O}(\ln(n)/\epsilon^2)\}$ columns, for some error parameter $\epsilon\in(0,1)$, and has no special structure.

\subsection{Leverage scores for rank deficient matrices}

In data science, matrices are likely to be rank deficient or close to rank deficient\footnote{As noted in \cite{udell2019big}, this is quite likely as the matrix becomes very large.}. Consequently, it is appropriate that algorithms take this into account; cf. the empirical evidence in \cite{pmlr-v95-chin18a}. 
{When the matrix has exactly rank$(A)=k\leq d$, then we can select a subset of exactly $k$ linearly independent columns of $A$ and the leverage scores of the subset are equivalent to those of $A$; this approach has been used in \cite{clarkson2013low_jacm,nelson2013osnap}. On the other hand, when the matrix is approximately low rank, then we must carefully choose a good subset of columns. In such cases it is common practice to consider the leverage scores of $A_k$; cf. \cite{drineas2012fast,gittens2013revisiting,perry2016augmented}.}

In Theorem \ref{thm:ls_qr_vs_svd}, we show that that if we 
{pick a set $\mathcal{K}\subseteq [d],|\mathcal{K}|=k$ appropriately, 
}
the leverage scores of $A_{:,\mathcal{K}}$ will be close to those of $A_k$. The approximation bound depends on the spectral gap of the original matrix: in particular, if the gap is large then the two leverage score distributions are identical or almost identical. If the spectral gap is small, however, they can differ substantially. We illustrate this experimentally in Section \ref{sec:experiments}. 

\begin{lemma}
\label{lem:subspace_distances}
Let $Q=\begin{pmatrix} Q_1 &  Q_2\end{pmatrix}$ and $U=\begin{pmatrix} U_1 &  U_2\end{pmatrix}$ be two $n\times n$ orthogonal matrices, such that $U_1$ and $Q_1$ have size $n\times k$, for some $k\in [n]$. The distance between range$(Q_1)$ and range$(U_1U_1^\top Q_1)$, which is the subspace spanned by its orthogonal projection on $U_1$, is equal to the distance between range$(U_1)$ and range$(Q_1Q_1^\top U_1)$, that is
\begin{align*}
    \|Q_1-U_1U_1^\top Q_1\| = \|U_1-Q_1Q_1^\top U_1\|.
\end{align*}
\begin{proof}
Take
\begin{align*}
    \|Q_1-U_1U_1^\top Q_1\| = \|U^\top(Q_1 - U_1U_1^\top Q_1)\| =
    \left\|
        \begin{pmatrix}
            U_1^\top Q_1\\
            U_2^\top Q_1            
        \end{pmatrix}
        -
        \begin{pmatrix}
            U_1^\top Q_1\\
            0_{(n-k)\times k}            
        \end{pmatrix}
    \right\|
    =
    \left\|
    \begin{pmatrix}
        0_{k\times k}\\
        U_2^\top Q_1
    \end{pmatrix}
    \right\|
\end{align*}
and
\begin{align*}
    \|U_1-Q_1Q_1^\top U_1\| = \|Q^\top(U_1 - Q_1Q_1^\top U_1)\| =
    \left\|
        \begin{pmatrix}
            Q_1^\top U_1\\
            Q_2^\top U_1            
        \end{pmatrix}
        -
        \begin{pmatrix}
            Q_1^\top U_1\\
            0_{(n-k)\times k}            
        \end{pmatrix}
    \right\|
    =
    \left\|
    \begin{pmatrix}
        0_{k\times k}\\
        Q_2^\top U_1
    \end{pmatrix}
    \right\|.
\end{align*}
From \cite[Thm 2.5.1]{golub2013matrix}, $\| U_2^\top Q_1\|=\|Q_2^\top U_1\|$, which concludes the proof.
\end{proof}
\end{lemma}

We can now state the main result.
\begin{theorem}
\label{thm:ls_qr_vs_svd}
Let $AP=QR$ be any pivoted QR of $A$, $AP=U\Sigma V^\top$ be the economy SVD of $AP$ and $1\leq k { \leq \text{rank}(A) }\leq d$ such that
\begin{align*}
AP=QR=\begin{pmatrix} Q_1 & Q_2 \end{pmatrix}
\begin{pmatrix} R_{11} & R_{12} \\ & R_{22} \end{pmatrix} \text { and }
AP=U\Sigma V^\top=U_1\Sigma_1V_1^\top + U_2\Sigma_2V_2^\top =A_k+E
\end{align*}
where $R_{11}, \Sigma_1$ are invertible and have size $k\times k$, $P$ is a $d\times d$ permutation matrix, $A_k=U_1\Sigma_1V_1^\top$ and $E=U_2\Sigma_2V_2^\top$. Let also $A_{:,\mathcal{K}}$ be the first $k$ columns of $AP$. The following holds
\begin{align*}
|\theta_i(A_k) - \theta_i(A_{:,\mathcal{K}})|
\leq
\left(\sqrt{\theta_i(A_k)} + \sqrt{\theta_i(A_{:,\mathcal{K}})}
\right)
\dfrac{\sigma_{k+1}(A)}{\sigma_{k}(R_{11})}.
\end{align*}

\begin{proof}
Let $\tilde V_1^\top, \tilde V_2^\top$ be the first $k$ columns of $V_1^\top$ and $V_2^\top$ accordingly. Since $R_{11}$ 
is invertible by assumption, the following {holds
\begin{align*}
Q_1R_{11} = U_1\Sigma_1\tilde V_1^\top + U_2\Sigma_2\tilde V_2^\top \Leftrightarrow
Q_1 = \left(U_1\Sigma_1\tilde V_1^\top + U_2\Sigma_2\tilde V_2^\top\right) R_{11}^{-1}.
\end{align*}
We then bound the distance of $U_1$ from its orthogonal projection on $Q_1$.
\begin{align*}
\|Q_1-U_1U_1^\top Q_1\| 
&= 
\left\|
    \left(
        U_1\Sigma_1\tilde V_1^\top + U_2\Sigma_2\tilde V_2^\top  
    \right) 
    R_{11}^{-1}
    - 
    U_1U_1^\top 
    \left(
        U_1\Sigma_1\tilde V_1^\top + U_2\Sigma_2\tilde V_2^\top 
    \right)
    R_{11}^{-1} 
\right\|
\\
&=
\left\|
    \left(
        U_1\Sigma_1\tilde V_1^\top 
        + 
        U_2\Sigma_2\tilde V_2^\top 
    \right)
    R_{11}^{-1} 
    - 
    U_1
    \left(
         U_1^\top U_1\Sigma_1\tilde V_1^\top 
        + 
        U_1^\top U_2\Sigma_2\tilde V_2^\top
    \right)
    R_{11}^{-1}  
\right\| 
\\
&=
\left\|
    U_1\Sigma_1\tilde V_1^\top R_{11}^{-1} 
    + 
    U_2\Sigma_2\tilde V_2^\top R_{11}^{-1} 
    - 
    U_1\Sigma_1\tilde V_1^\top R_{11}^{-1} 
\right\|
\\
&=
\left\|
    U_2\Sigma_2\tilde V_2^\top R_{11}^{-1}  
\right\|
\leq
\|\Sigma_2\|\|R_{11}^{-1}\|=\dfrac{\sigma_{k+1}(A)}{\sigma_{k}(R_{11})}.
\end{align*}}
From Lemma \ref{lem:subspace_distances}, the same bound also holds for
$\|Q_1-U_1U_1^\top Q_1\|$.

Finally take the absolute difference of the leverage scores
\begin{align*}
|\theta_i(Q_1) - \theta_i(U_1)| &= \left| e_i^\top U_1U_1^\top e_i - e_i^\top Q_1Q_1^\top e_i \right|\\
&= \left| e_i^\top U_1U_1^\top e_i - e_i^\top U_1U_1^\top Q_1Q_1^\top e_i - e_i^\top(I-U_1U_1^\top)Q_1Q_1^\top e_i \right|\\
&\leq \left| e_i^\top U_1(U_1^\top  - U_1^\top Q_1Q_1^\top) e_i\right| +\left|e_i^\top(Q_1-U_1U_1^\top Q_1)Q_1^\top e_i \right|\\
&\leq \| e_i^\top U_1\|\|(U_1^\top  - U_1^\top Q_1Q_1^\top)\| +\|(Q_1-U_1U_1^\top Q_1)\|\|Q_1^\top e_i \| \\
&\leq \left(\| e_i^\top U_1\| +\|Q_1^\top e_i \|\right)\dfrac{\sigma_{k+1}(A)}{\sigma_{k}(R_{11})} 
=
\left(\sqrt{\theta_i(U_1)} + \sqrt{\theta_i(Q_1)}\right)
\dfrac{\sigma_{k+1}(A)}{\sigma_{k}(R_{11})} .
\end{align*}
\end{proof}
\end{theorem}

{From Theorem \ref{thm:ls_qr_vs_svd} it is evident that if we choose $k=\text{rank}(A)$ then $\sigma_{k+1}(A)=0$ and the leverage scores of $Q_1$ and $U_1$ are equivalent.}

\section{Tools from numerical linear algebra\label{sec:tools}} 

We review the basic tools used in the proposed methods, namely sketching with normal random projections, maximal linearly independent column selection and rank revealing QR factorizations.

\subsection{Rank estimation and column subset selection}
{
From Theorem \ref{thm:ls_qr_vs_svd} it is evident that a matrix consisting of a maximal set of linearly independent columns of $A$ will have the same leverage scores as $A$. In \cite{clarkson2013low_jacm,nelson2013osnap} such a set of columns is obtained by using the state-of-the-art algorithms of Cheung et al \cite[Theorems 2.7, 2.11]{cheung2013fast} for this task. We state a result regarding the complexity of these algorithms for tall-and-thin matrices in Corollary \ref{cor:cheung_complexity_tall_and_thin}. Essentially, this is a slight modification of \cite[Theorem 2.11]{cheung2013fast} where we remove the $\log d$ factor from ${\tt nnz}(A)$ at the expense of changing the factor $k^\omega\log d$ to $d^\omega$, where $k=$rank$(A)\leq d$.
\begin{corollary}
    There exists an algorithm such that, for any matrix $A\in\mathbb{R}^{n\times d}$ with $n>d$, it can find a set of $\text{rank}(A)$ linearly independent columns in $\mathcal{O}({\tt nnz}(A) + d^\omega)$ operations with probability at least $1-\mathcal{O}(d^{-1/3})$.
    \begin{proof}
        We use the procedure described in the proof of \cite[Theorem 2.11]{cheung2013fast}. Specifically, we apply first \cite[Lemma 2.9]{cheung2013fast} with $k=\mathcal{O}(d)$ to reduce $A$ to a matrix $\hat A$ of size $\mathcal{O}(d)\times d$ in $\mathcal{O}({\tt nnz}(A))$, and we skip the repetitive application of \cite[Lemma 2.10]{cheung2013fast}. We then find a set of linearly independent columns  with Gaussian elimination in $\mathcal{O}(d^\omega)$ directly on $\hat A$.
    \end{proof}
\label{cor:cheung_complexity_tall_and_thin}
\end{corollary}
}

\subsection{Rank revealing QR}

{Recall that the bound of Theorem \ref{thm:ls_qr_vs_svd} depends on the smallest singular value of $1/\sigma_k(R_{11})=1/\sigma_k(A_{:,\mathcal{K}})$. Therefore, to minimize the approximation error, we need to choose $A_{:,\mathcal{K}}$ such that $\|A_{:,\mathcal{K}}^\dagger\|$ is minimized. For this task we deploy a rank revealing QR method (RRQR).} The literature on RRQR is extensive \cite{bischof1998computing,businger1965linear,chandrasekaran1994rank,demmel2015communication,quintana1998blas}, and in this section we focus on the \textit{strong} RRQR (SRRQR) as defined by Gu and Eisenstat \cite{gu1996efficient}.

\begin{definition}
Given a matrix $A$ of size $n\times d$ with $n\geq d$, a SRRQR factorization of $A$ is a pivoted QR factorization of the form \begin{align*}
    AP = QR = \begin{pmatrix} {Q}_1 & {Q}_2\end{pmatrix} \begin{pmatrix} {R}^{(k)}_{11} & {R}^{(k)}_{12} \\ & {R}^{(k)}_{22} \end{pmatrix}
\end{align*}
such that ${R}^{(k)}_{11}$ has size $k\times k$,
$P$ is a permutation matrix and ${R}^{(k)}_{11}$ and ${R}^{(k)}_{22}$ are upper triangular and the following inequalities are satisfied
\begin{align}
    \sigma_i({R}^{(k)}_{11})
    \geq 
    \dfrac{\sigma_i(A)}{p_1(k,n)} 
    \text{ and }
    \sigma_j({R}^{(k)}_{22})
    &\leq 
    \sigma_{k+j}(A)p_1(k,n)\\
    \text{ and also }
    \left|
        \left(
            ({R}^{(k)}_{11})^{-1}
            {R}^{(k)}_{12}
        \right)_{i,j}
    \right|
    &\leq p_2(k,n)
\end{align}
for $i\in[k]$ and $j\in[n-k]$, where $p_1(k,n)$ and $p_2(k,n)$ are functions bounded by low-degree polynomials in $k$ and $n$.
\label{def:srrqr}
\end{definition}

A strong RRQR factorization can be computed with \cite[Algorithm 5]{gu1996efficient}, which, given a parameter $ \phi\geq 1$, has the following properties.
\begin{align}
    \label{eq:srrqr_lower}
    \sigma_i({R}^{(k)}_{11}) 
    &\geq
    \dfrac{\sigma_i(A)}{\sqrt{1+ \phi^2k(d-k)}}, i\in[k]\\
    \text{ and }
    \sigma_j({R}^{(k)}_{22})
    &\leq
    \sigma_{j+k}(A)\sqrt{1+ \phi^2k(d-k)}, j\in[n-k].
    \label{eq:srrqr_upper}
\end{align}
The number of operations required for this algorithm is bounded by $\mathcal{O}(ndk\log_\phi\sqrt{d})$. For ease of reference, we list an abstract description in Algorithm \ref{alg:srrqr}. For details we defer the reader to the original paper. 

\begin{algorithm}[htb]
    \caption{SRRQR$(A, \zeta, \phi)$. Strong RRQR factorization \cite{gu1996efficient}.}
    \begin{algorithmic}[1]
        \Require matrix $A \in \mathbb{R}^{n \times d}$, $n\geq d$, tolerance for small singular values $\zeta$, parameter $\phi\geq 1$.
		\Ensure $P,Q,R$ such that $AP=QR$, satisfying bounds \ref{eq:srrqr_upper} and \ref{eq:srrqr_lower}.
        \State Run Algorithm $5$ of \cite{gu1996efficient} on $A$ with parameters $\phi, \zeta$, detecting the rank $k$ of $A$ as $k=\arg\max_{j\in[d-1]}\left\{\sigma_j(A) \middle\vert \frac{\sigma_{j+1}(A)}{\sigma_j(A)}\leq \zeta\sqrt{j(d-j)}\right\}$.
        \State \Return $P,Q,R,k$
    \end{algorithmic}
    \label{alg:srrqr}
\end{algorithm}

Algorithm \ref{alg:srrqr} detects the presence of a suitably large gap in the singular values of $A$ (\cite{gu1996efficient}, Remark 1). More specifically, given a tolerance $\zeta$, it can detect whether the following holds for the singular value gap at step $k$
\begin{align*} 
\dfrac{\zeta}{1+2\phi^2(n-k)}
\leq 
\dfrac{\sigma_{k+1}(A)}{\sigma_{k}(A)}
\leq  
\zeta \sqrt{k(n-k)}.
\end{align*}

\subsection{Sketching\label{sec:tools_sketching}}
We recall two useful tools from the RandNLA literature, namely Johnson-Lindenstrauss Transforms (JLT) and Oblivious Subspace Embeddings (OSE).
The Johnson-Lindenstrauss lemma (\cite{johnson1984extensions}) states that a set of $n$ vectors in $d$ dimensions can be mapped down to ${\cal O}(\log n)$ dimensions while their pairwise dot products are approximately preserved up to some multiplicative constant. Of interest is the case when $S$ is a matrix drawn from a random distribution, for which it holds that $S$ will be an $\epsilon$-JLT for a fixed set of points with probability at least $1-\delta$, for some small failure parameter $\delta\in(0,1)$. 
\begin{definition}
	A random matrix $S \in \mathbb{R}^{r\times d}$ is a Johnson-Lindenstrauss transform (JLT) with parameters $\epsilon,\delta,n$, or
	$(\epsilon,\delta)$-JLT for short, if with probability at least $1-\delta$, for any $n$-element set $V \subseteq \mathbb{R}^{d}$ it holds that 
$|(S{ v})^\top Sw-v^\top w |\leq\epsilon \| v\|\| w\|$ for any $ v, w \in V$.
	\label{def:JLT}
\end{definition}
Choosing $ v= w$, it follows that the JLT approximately preserves vector lengths as well.
The use of
JLT was proposed in \cite{indyk1998approximate} where it was constructed as an $r\times n$ matrix with independent standard normal random variables as elements;
cf. \cite{achlioptas2001database,ailon2006approximate,ailon2009fast,clarkson2013low_jacm,dasgupta2010sparse,kane2014sparser,meng2013low,nelson2013sparsity,sarlos2006improved,tropp2011improved} for further discussion and  refinements such as obtaining a sparser or structured $S$ for faster multiplications. Oblivious Subspace Embeddings generalize this definition for an entire subspace rather than a finite set of vectors; cf.
\cite{sarlos2006improved}. As with JLT, we work with random matrices which are OSE for a fixed subspace with high probability.
\begin{definition}
	A random matrix $S \in \mathbb{R}^{r\times n}$ is an $(\epsilon,\delta)$-Oblivious Subspace Embedding ($(\epsilon,\delta)$-OSE) with parameters $\epsilon,\delta\in(0,1)$, if with probability at least $1-\delta$, for any fixed matrix $A\in\mathbb{R}^{n\times d}$, $n>d$, it holds that 
\begin{equation}
\label{eq:OSE_property}
(1-\epsilon)\| x\|\leq \|S x\| \leq (1+\epsilon)\| x\|
\end{equation} for any $ x \in$ range$(A)$ with probability at least $1-\delta$.
\label{def:OSE}
\end{definition}
We next review some constructions for JLTs and OSEs.

\subsection{Constructing randomized embeddings}
{From the extensive literature on the subject, (cf. the surveys  \cite{bourgain2015toward,kannan2017randomized,mahoney2012randomized,martinsson2020randomized,woodruff2014sketching}), we recall some well known methods for constructing randomized embeddings and their properties that are essential in the sequel. We use $S$ to denote an $r \times n$ randomized embedding and assume that $n>r$. }

\subsubsection{Gaussian embeddings \cite{indyk1998approximate}}
In this case $S=\frac{1}{\sqrt{r}}G$, where $G$ is a $r\times n$ Gaussian matrix. 
\begin{lemma}\label{def:GJLT}
Let $G\in\mathbb{R}^{d\times r}$ be a Gaussian matrix, rescaled by $1/\sqrt{r}$.
For a fixed set $V$ of $n$ vectors in $\mathbb{R}^d$, $G$ is an $(\epsilon,\delta)$-JLT for $V$ if $r\geq \dfrac{4\log n}{\epsilon^2/2-\epsilon^3/3}$, with $\delta\leq 1/n$.\\
\end{lemma}

Gaussian embeddings can also provide OSEs. The following lemma will be used later to construct preconditioners.

\begin{lemma}
    \label{lem:sdgu}
    \cite{davidson2001local,meng2014lsrn}
    The following inequality holds for minimum singular value of the matrix $GU\in\mathbb{R}^{m\times k}, m>k$, where $G\in\mathbb{R}^{m\times n}$ is a Gaussian matrix and $U\in\mathbb{R}^{n\times k}$ is a matrix with orthonormal columns. Then for any $\alpha\in(0,1-\sqrt{k/m})$
    \begin{align*}
    \min
    \left\{
        \mathbb{P}
        \left[
            \sigma_1(GU) < (1+\alpha)\sqrt{m}+\sqrt{k}
        \right],
        \mathbb{P}
        \left[
            \sigma_k(GU) > (1-\alpha)\sqrt{m}-\sqrt{k}
        \right]
    \right\}
    \geq 1 - e^{-\alpha^2m/2}.
    \end{align*}
\end{lemma}
{
Lemma \ref{lem:sdgu} can be extended for matrices $A$ whose columns are not orthonormal.}

\begin{corollary}
\label{cor:sdga}
{Let $A\in\mathbb{R}^{n\times d}$, where $n>d$ and rank$(A)=k\leq d$. Then for any $\alpha\in(0,1-\sqrt{k/m})$
\begin{align*}
\min
&
\left\{
\mathbb{P}
    \left[
        \sigma_1(GA) < \sigma_1(A)
        \left(
            (1+\alpha)\sqrt{m}+\sqrt{k}
        \right)
    \right],
\right.
\\
&\ \:
\left.
    \mathbb{P}
    \left[
        \sigma_k(GA) > \sigma_k(A)
        \left(
            (1-\alpha)\sqrt{m}-\sqrt{k}
        \right)
    \right]
\right\}
\geq 1 - e^{-\alpha^2m/2}.
\end{align*}}

{
\begin{proof}
Let $A=U\Sigma V^\top$ be the economy SVD of $A$, and let $G\in\mathbb{R}^{m\times n}$ be a Gaussian matrix where $m>d\geq k$. Lemma \ref{lem:sdgu} suggests that, with probability at least $1-2e^{-\alpha^2m/2}$, both events hold at the same time and therefore for any vector $y\in\text{range}(U)$ it holds that 
\begin{align*}
    \|y\|\left((1-\alpha)\sqrt{m}-\sqrt{k} \right)
    < 
    \|Gy\|
    <
    \|y\|\left((1+\alpha)\sqrt{m}+\sqrt{k}\right).
\end{align*}
Since $\text{range}(A)=\text{range}(U)$, the same holds for any vector $y\in\text{range}(A)$. Therefore for any nonzero $x\in R^d$, we can write
\begin{align*}
    \|Ax\|\left((1-\alpha)\sqrt{m}-\sqrt{k} \right)
    < 
    \|GAx\|
    <
    \|Ax\|\left((1+\alpha)\sqrt{m}+\sqrt{k}\right).
\end{align*}
Since for any such vector $x$ it holds that $\sigma_k(A)\leq \frac{\|Ax\|}{\|x\|}\leq \sigma_1(A)$, combining the above two inequalities it follows that
\begin{align*}
\sigma_k(GA) > \sigma_k(A)\left((1-\alpha)\sqrt{m}-\sqrt{k} \right) \text{ and } \sigma_1(GA) < \sigma_1(A)\left((1+\alpha)\sqrt{m}+\sqrt{k}\right).
\end{align*}
\end{proof}
}
\end{corollary}

\subsubsection{Subsampled Randomized Hadamard Transform (SRHT)  \cite{ailon2006approximate,ailon2009fast}}
We follow the analysis of \cite{tropp2011improved}.
The SRHT are constructed as matrices of the form $S=\sqrt{\frac{n}{r}}PHD$, where 
$D$ is a diagonal matrix with diagonal elements equal to $\pm 1 $ with probability $1/2$, 
$H$ is a $n\times n$ Walsh-Hadamard matrix rescaled by $1/\sqrt{n}$ and 
$P$ is a $r\times n$ matrix which samples $r$ out of $n$ coordinates of a vector uniformly at random. 
In this case we assume that $n$ is a power of $2$.

\begin{theorem}\cite[Thm. 3.1]{tropp2011improved}\label{def:FOSE}
For a fixed matrix $U\in\mathbb{R}^{n\times d}$ with orthonormal columns, if $F\in\mathbb{R}^{r\times n}$ is a SRHT embedding and if $r\geq 4\left[\sqrt{d}+\sqrt{\log (nd)}\right]^2\log d$, then
\begin{align*}\sqrt{1/6}\leq \sigma_d(FU) \leq \sigma_1(FU) \leq \sqrt{13/6}\end{align*}
holds with probability at least $1-3/d$.
Moreover, the product $PHD x$ for a vector $ x$ can be computed in $O(n\log r)$ arithmetic operations.
\end{theorem}

\subsubsection{CountSketch \cite{charikar2002finding,clarkson2013low_jacm,meng2013low,nelson2013osnap}}
In this case the corresponding matrices have exactly one non-zero element in each column at a random position chosen uniformly, and each nonzero is equal to $\pm 1$ with probability $1/2$. {More formally, the matrix $S$ is specified by a hash function $h: [n]\rightarrow[r]$, such that $\forall i\in[n], h(i)=r_i$ for $r_i\in[r]$ with probability $1/r$. We then set each element $S_{h(i),i}=s_i$, $\forall i\in[n]$, where $s_i=\pm 1$ with probability $1/2$. The remaining the elements of $S$ are set to zero}.

\begin{theorem}\cite[Thm. 3]{ nelson2013osnap}\label{thm:SOSE}
If $S\in\mathbb{R}^{r\times n}$ is a CountSketch with $r\geq \dfrac{d^2+d}{\delta(2\epsilon-\epsilon^2)^2}$ and $U\in\mathbb{R}^{n\times d}$ is a fixed matrix with orthonormal columns, then $S$ is an $(\epsilon,\delta)$-OSE for $U$. Moreover, the product $SA$ for a sparse matrix $A$ can be computed in $O({\tt nnz}(A))$, the random signs need to be 4-wise independent while the hash function $h$ needs to be 2-wise independent.
\end{theorem}

\section{Compositions of randomized embeddings\label{sec:compositions}}

The main drawback of Gaussian embeddings is that the product $GA$ costs $O({\tt nnz}(A)d)$ arithmetic operations and also requires the generation of $mn$ random numbers from $\mathcal{N}(0,1)$, which is expensive. Sparse embeddings can be applied faster, but to achieve similar distortion, more samples are required.

The possibility of combining sparse and dense transforms to further reduce the size of the sketched matrix {is not new and to the best of our knowledge, has been first mentioned briefly in \cite{clarkson2013low_jacm}}.
We extensively use the so called ``CountGauss'' transform \cite{kapralov2016fake}, which is very efficient in  practice and is of the form $\tilde G=GS$, where $G\in\mathbb{R}^{m\times r}$ 
is a Gaussian matrix and $S\in\mathbb{R}^{r\times n}$ is a CountSketch. 
{In Algorithm \ref{alg:count_gauss_sketch} we describe the sketching procedure. The properties are detailed in the rest of the section.}

\begin{algorithm}[htb]
\caption{CountGaussSketch$(A,\gamma,\delta,\epsilon)$: Sketching very tall sparse matrices with CountGauss embeddings.}
\label{alg:count_gauss_sketch}
\begin{algorithmic}[1]
	\Require Matrix $A\in \mathbb{R}^{n\times d}$ with $n\gg \mathcal{O}(d^2)$, ${\tt nnz}(A)>\mathcal{O}(d^3)$, oversampling factor $\gamma \in (1,3]$, CountSketch OSE parameters $\epsilon,\delta\in(0,1)$.
	\Ensure Sketched matrix $\tilde A = GSA$.
	\State Set $r=\left\lceil\dfrac{d^2+d}{(2\epsilon-\epsilon^2)^2\delta}\right\rceil$. \Comment{A practical choice is $\epsilon=\frac{1}{2},\delta=\frac{1}{3} \Rightarrow r\approx 5(d^2+d)$.}
	\State Set $m=\lceil \gamma d\rceil$. \Comment{A practical choice is $\gamma=2$.}
	\State Compute $\tilde A = G(SA)$ where $G$ is a $m\times r$ Gaussian matrix and $S$ is a $r\times n$ CountSketch. \Comment{$\mathcal{O}({\tt nnz}(A)+mdr)$.}
	\State \Return $\tilde A$.
\end{algorithmic}
\end{algorithm}

\subsection{Sketching cost\label{sec:sketching_cost_analysis}}
Selecting $S$ to be a CountSketch OSE, the complexity of the product $\tilde GA$ is $\mathcal{O}({\tt nnz} (A))$ to compute $SA$, and then $\mathcal{O}(mdr)$ to compute $G(SA)$. 
The cost can be further reduced if we replace $S$ with $\tilde S=FS$, where $F\in\mathbb{R}^{t\times r}$ is a SRHT and $S\in\mathbb{R}^{r\times n}$ is, again, a CountSketch. Then $\tilde S$ is a $(\tilde \epsilon, \tilde \delta)$-OSE, with $\tilde \epsilon\approx2\epsilon$ and $\tilde \delta \approx 2\delta$, while the product $F(SA)$ can be computed in $\mathcal{O}({\tt nnz}(A)+rd\log t)$, which is roughly $\mathcal{O}({\tt nnz}(A)+d^3\log (d\log d))$, since we need $\mathcal{O}({\tt nnz}(A))$ for $SA$ and $\mathcal{O}(rd\log t)$ for $F(SA)$.
Note that in order for $\tilde G A$ to be beneficial in comparison to $GA$, the original matrix needs to satisfy $n\gg d^2$ and ${\tt nnz}(A)\gg d^3$.

\subsection{Spectrum approximation}
It is worth noting that using Algorithm \ref{alg:count_gauss_sketch} the spectrum of $\tilde A$ is approximately the same with the spectrum of $A$.

\begin{lemma}
\label{lem:spectrum_gsu}
Given $U\in \mathbb{R}^{n\times k}$, a matrix with orthonormal columns, if we use Algorithm \ref{alg:count_gauss_sketch} on $U$ to obtain $\tilde U=GSU$, the following hold with probability at least $(1-\delta)(1-2e^{-\alpha^2m/2})$.
\begin{align*}
    \sigma_1(GSU) < ((1+\alpha)\sqrt{m}+\sqrt{k})(1+\epsilon)
    \text{\ \ and\ \ }
    \sigma_{k}(GSU) > ((1-\alpha)\sqrt{m}-\sqrt{k})(1-\epsilon).
\end{align*}
\begin{proof}
From Definition \ref{def:OSE} and Corollary \ref{cor:sdga}, we have that
    \begin{equation*}
        \sigma_1(GSU)=\|GSU\|<((1+\alpha)\sqrt{m}+\sqrt{k})\|SU\|\leq ((1+\alpha)\sqrt{m}+\sqrt{k})(1+\epsilon)
    \end{equation*}
    and also
    \begin{equation*}
        \sigma_{k}(GSU)>((1-\alpha)\sqrt{m}-\sqrt{k})\sigma_{k}(SU)\geq ((1-\alpha)\sqrt{m}-\sqrt{k})(1-\epsilon)
    \end{equation*}
    both hold with probability at least $(1-\delta)(1-2e^{-\alpha^2m/2})$.
\end{proof}
\end{lemma}

\begin{lemma}
\label{lem:sigma_min_gsa}
{
Given $A\in \mathbb{R}^{n\times d}$, if we use Algorithm \ref{alg:count_gauss_sketch} on $A$ to obtain $\tilde A=GSA$, for any $k\leq\text{rank}(A)\leq d$ the following holds with probability at least $(1-\delta)(1-2e^{-\alpha^2m/2})$
\begin{align*}
    \sigma_k(GSA) > ((1-\alpha)\sqrt{m}-\sqrt{k})(1+\epsilon)\sigma_k(A).
\end{align*}
\begin{proof}
From the definition of the singular values, we have that $\sigma_k(GSA)=\frac{\|GSAx^*\|}{\|x^*\|}$, for some $x^*=Ux'$ where $x'\in\mathbb{R}^k$ and $U\in\mathbb{R}^{d\times k}$ has orthonormal columns. From Definition \ref{def:OSE} and Corollary \ref{cor:sdga} we have that for any $x\in\mathbb{R}^{k}$ it holds that 
\begin{align*}
\|GS(AU)x\|
>
((1-\alpha)\sqrt{m}-\sqrt{k})\|S(AU)x\|
\geq
((1-\alpha)\sqrt{m}-\sqrt{k})(1-\epsilon)\|(AU)x\|.
\end{align*}
Therefore, since $x^*=Ux'$ belongs to a $k$-dimensional subspace we have that
\begin{align*}
\sigma_k(GSA)
=
\frac{\|GSAx^*\|}{\|x^*\|}
>
((1-\alpha)\sqrt{m}-\sqrt{k})(1-\epsilon)\frac{\|Ax^*\|}{\|x^*\|}
\geq
((1-\alpha)\sqrt{m}-\sqrt{k})(1-\epsilon) \sigma_k(A)
\end{align*}
holds with probability at least $(1-\delta)(1-2e^{-\alpha^2m/2})$.
\end{proof}
}
\end{lemma}

\subsection{Column subset selection}
We next combine Algorithm \ref{alg:count_gauss_sketch} with a SRRQR factorization 
and show that it can be used as an efficient preprocessing step for 
leverage scores computations. We list this in Algorithm \ref{alg:count_gauss_strong_rrqr}. {The complexity of this algorithm is $\mathcal{O}({\tt nnz}(A)+\gamma d^4/(\epsilon^2\delta))$ to compute $\tilde A = GSA$ plus $\mathcal{O}(\gamma d^2k\log_\phi\sqrt{d})$ to run SRRQR on $\tilde A$.}

\begin{algorithm}[htb]

\caption{CountGaussSRRQR$(A,\gamma,\delta,\epsilon,\zeta,\phi)$: Column subset selection of tall-and-thin matrices with CountGauss embeddings and Strong RRQR.}
\label{alg:count_gauss_strong_rrqr}
\begin{algorithmic}[1]
	\Require Matrix $A\in \mathbb{R}^{n\times d}$ with $n\gg \mathcal{O}(d^2)$, ${\tt nnz}(A)>\mathcal{O}(d^3)$, oversampling factor $\gamma \in (1,3]$, CountSketch OSE parameters $\epsilon,\delta\in(0,1)$, tolerance $\zeta$ for small singular values, SRRQR parameter $\phi>1$.
	\Ensure Estimation $k$ for \text{rank}$(A)$, column permutation matrix $P \in \mathbb{R}^{d\times k}$, matrix $R \in \mathbb{R}^{k\times k}$.
	\State {Compute $\tilde A\leftarrow $CountGaussSketch$(A, \gamma, \delta, \epsilon)$. \Comment{$\mathcal{O}({\tt nnz}(A)+\gamma d^4/(\epsilon^2\delta)).$}}
	\State Compute $P,Q,R,k\leftarrow$SRRQR$(\tilde A,\zeta,\phi)$ such that $\tilde AP= Q R$. \Comment{$\mathcal{O}(\gamma d^2k\log_\phi\sqrt{d})$.}
	\State \Return $P_{:,1:k}, R_{1:k,1:k}, k$.
\end{algorithmic}

\end{algorithm}

Algorithm \ref{alg:count_gauss_strong_rrqr} essentially runs SRRQR on $\tilde A = GSA$ to obtain a column sampling matrix {$\tilde P_k\in\mathbb{R}^{d\times k}$} such that the approximation bounds \ref{eq:srrqr_lower} and \ref{eq:srrqr_upper} are satisfied for $\tilde A$. 
We next derive a lower bound for the smallest singular value of $A\tilde P_k$, which consists of the ``best'' $k$ columns of $A$ according to the aforementioned permutation $\tilde P_k$ for $GSA$.

\begin{lemma}
    \label{lem:bound_sigma_min_r_11}
    Assume we get $\tilde P_k\in\mathbb{R}^{d\times k},\tilde R_k\in\mathbb{R}^{k\times k},k$ by running Algorithm \ref{alg:count_gauss_strong_rrqr} on $A$. Let
    \begin{align*}A\tilde P_k=Q_kR_k
    \text { and }
    GSA\tilde P_k=
    \tilde Q_k\tilde R_k
    \end{align*}
    where $Q_k\in\mathbb{R}^{n\times k},R_k\in\mathbb{R}^{k\times k}$ are taken from any QR factorization of $A\tilde P_k$, and $R_{k}$,$\tilde R_{k}$ have full rank $k$.
    Then with probability at least $(1-\delta)(1-2e^{-\alpha^2m/2})$ the following holds
    \begin{align*}
        \sigma_{k}(R_k) 
        > 
        \dfrac{\sigma_k(A)}{\xi\eta \rho}
    \end{align*}
    where $\xi = \frac{(1+\alpha+\sqrt{k/m})}{(1-\alpha-\sqrt{k/m})}$, $\eta=\frac{(1+\epsilon)}{(1-\epsilon)}$ and $\rho=\sqrt{1+\phi^2k(d-k)}$.
    \begin{proof}
    {
    We can write $GSQ_kR_k=\tilde Q_k \tilde R_k$. Since $\tilde R_k$ is full column rank $k$ and $\tilde Q_k$ has orthonormal columns, then both $R_k$ and $GSQ_k$ have full column rank as well. With this observation, we can work as follows
    \begin{align*}
        GSQ_kR_k
        =
         \tilde Q_k\tilde R_k 
        \Rightarrow 
        & \tilde Q_k^\top GSQ_k R_k
        =
        \tilde R_k 
        \Leftrightarrow
        (\tilde Q_k^\top GSQ_k R_k)^{-1}
        =
        \tilde R_k^{-1} 
        \Leftrightarrow 
        \\
        & R_k^{-1} (\tilde Q_k^\top GSQ_k )^{-1}
        =
        \tilde R_k^{-1}
        \Leftrightarrow 
        R_k^{-1}
        = 
        \tilde R_k^{-1} (\tilde Q_k^\top GSQ_k ).
    \end{align*}}
    We can then bound the $k$-th singular value of $R_k$ from below as follows
    {\begin{align*} 
        \|R_k^{-1}\| 
        &= 
        \| \tilde R_k^{-1} \tilde Q_k^\top GSQ_k  \|
        \leq \| \tilde R_k^{-1}\| \| GSQ_k  \|
        \Leftrightarrow
        \sigma_k(R_k) \geq \frac{\sigma_k(\tilde R_k)}{\sigma_1(GSQ_k)}
    \end{align*}
    }
    and also, from Inequality \ref{eq:srrqr_lower} we have that
    $
        \sigma_i (\tilde R_k) 
        \geq
        \dfrac{\sigma_i(GSA)}{\sqrt{1+\phi^2k(d-k)}}.
    $
    Combining these two inequalities with 
    {Lemmas \ref{lem:spectrum_gsu} and \ref{lem:sigma_min_gsa}
    it follows that
    \begin{align*}
    \sigma_{k}(R_k)
    \geq
    \dfrac{\sigma_k(\tilde R_k)}{\sigma_1(GSQ_k)}
    \geq
    \dfrac{\sigma_k(GSA)}{\sigma_1(GSQ_k)\rho}
    >
    \dfrac{
        (1-\alpha-\sqrt{k/m})(1-\epsilon)
    }
    {
        (1+\alpha+\sqrt{k/m})(1+\epsilon)
    }
    \dfrac{\sigma_k(A)}{\rho}
    \end{align*}
    }
    holds with probability at least $(1-\delta)(1-2e^{-\alpha^2m/2})$, where $\rho=\sqrt{1+\phi^2k(d-k)}$.
    \end{proof}
\label{lem:good_p}
\end{lemma}

\section{Leverage scores algorithms\label{sec:ls_algorithms}}
We are now in a position to present our results and algorithms for our primary target, which is leverage scores computations of rectangular matrices.
Without loss of generality, we assume that each row and column of $A$ has at least one nonzero. 

\subsection{A leverage scores algorithm based on the hat matrix formulation}

We present a straightforward approach for computing \rsls\  based on Equation \ref{eq:ls_orthonormal_basis}, listed as Algorithm \ref{alg:ls_direct}, which is simple and practical. 
In terms of complexity, we first state the following.
\begin{lemma}
\label{lem:gram_complexity}
The Gram matrix $B=A^\top A$ can be computed in $\mathcal{O}({\tt nnz_2}(A))$ operations.
\begin{proof}
This is easy to verify from the sum of column-row outer products formulation of matrix multiplication, e.g.
\begin{equation*}
    B = A^\top A = A_{1,:}A_{1,:}^\top + A_{2,:}A_{2,:}^\top + ... + A_{n,:}A_{n,:}^\top.
\end{equation*}
{Each rank-1 term has $\mathcal{O}({\tt nnz}^2(A_{i,:}))$ nonzeros which can be computed and added to the final result in $\mathcal{O}({\tt nnz}^2(A_{i,:}))$ operations. }
\end{proof}
\end{lemma}

We can therefore compute $A^\top A$ in $\mathcal{O}({\tt nnz_2}(A))$ using Lemma \ref{lem:gram_complexity} and then compute the {SVD} of $(A^\top A)=V\Sigma^2 V^\top$ in $\mathcal{O}(d^3)$ and keep only the $k$ {singular values/vectors} in $\Sigma_k=\Sigma_{1:k,1:k}$ and $V_k=V_{1:d,1:k}$, where $k$ is the rank of the matrix. Finally, the leverage scores of $A_k$ can be returned as $\theta_i=\|e_i^\top AV_k\Sigma^{-1}_k\|^2$ in $\mathcal{O}({\tt nnz_2}(A))$ operations, and therefore the total complexity of the algorithm is $\mathcal{O}({\tt nnz_2}(A)+d^3)$. 

\begin{algorithm}[htb]
\caption{Leverage scores  computation {via Gram pseudoinverse}.}
\label{alg:ls_direct}
\begin{algorithmic}[1]
	\Require Matrix $A\in \mathbb{R}^{n\times d}$ with $n\gg d$ and rank $k\leq d$.
	\Ensure \rsls \ $\theta_i$ of $A$ over its dominant $k$-subspace.
	\State Compute $B\leftarrow A^\top A$ \Comment{$\mathcal{O}({\tt nnz_2}(A))$}
	\State Compute {the SVD of $B=V\Sigma^2V^\top$}. \Comment{$\mathcal{O}(d^3)$}
	\State Keep only the top $k$ non-zero {singular values} and the corresponding {singular vectors} in the matrices $\Sigma_k:=\Sigma_{1:k,1:k}$ and $V_k:=V_{1:d,1:k}$.
	\State \Return {$\theta_i=\|e_i^\top AV_k\Sigma_k^{-1}\|^2$}. \Comment{$\mathcal{O}({\tt nnz_2}(A))$.}
\end{algorithmic}
\end{algorithm}

\subsection{A sparse pivoted QR approach}
If $A$ is sparse, instead of computing the QR factorization by dense QR methods it is preferable to use a sparse QR \cite{davis2011algorithm}, or a ``Q-less'' QR method with column pivoting-based Gram-Schmidt  (e.g. \cite{berry2005algorithm,stewart1999four}). 
We adopt the SPQR method from \cite{berry2005algorithm} which returns the upper triangular $R_k\in\mathbb{R}^{k\times k}$ factor and a column permutation matrix $P_k\in\mathbb{R}^{d\times k}$, while  avoiding the explicit formation of $Q$. 
Assuming that, on average, each column of $A$ has $\mathcal{O}({\tt nnz}(A)/d)$ non-zeros, then the \textit{average} complexity of the SPQR method of \cite{berry2005algorithm} would be $\mathcal{O}({\tt nnz}(A)+{\tt nnz}(A)k^2/d)$. From $P_k,R_k$, we can obtain the \rsls\ of $AP_k$ by inverting $R_k$ in $\mathcal{O}(k^3)$ and computing the row norms
\begin{equation}
    \theta_i = \|e_i^\top AP_k R_k^{-1}\|_2^2
    \label{eq:ls_via_pqr}
\end{equation} in $\mathcal{O}({\tt nnz_2}(AP_k))$. We list this method as Algorithm \ref{alg:ls_spqr}.

\begin{algorithm}[htb]
\caption{Exact leverage scores with SPQR \cite{berry2005algorithm,stewart1999four}.}
\label{alg:ls_spqr}
\begin{algorithmic}[1]
	\Require Matrix $A\in \mathbb{R}^{n\times d}$ with $n\gg d$ and rank $k\leq d$.
	\Ensure \rsls \ $\theta_i$ of $A$ over its dominant $k$-subspace.
	\State Compute $P_k,R_k$ with ``Q-less'' SPQR. \Comment{$\mathcal{O}({\tt nnz}(A)+ {\tt nnz}(A)k^2/d)$ (average)}
	\State Set $A_{:,\mathcal{K}}\leftarrow AP_k$.
	\State \Return $\theta_i=\|e_i^\top A_{:,\mathcal{K}}R_k^{-1}\|^2$. \Comment{$\mathcal{O}({\tt nnz_2}(A_{:,\mathcal{K}}))$ }
\end{algorithmic}
\end{algorithm}

Algorithm \ref{alg:ls_spqr} is better suited for matrices with small rank $k\ll d$. For close to full rank matrices, Algorithm \ref{alg:ls_direct} will be faster. 
Even though SPRQ does not provide the strong RRQR guarantees, in general it works well in practice. 

\subsection{Algorithm LS-HRN-exact}
Based on these findings, we are ready to describe Algorithm \ref{alg:ls_hrn_exact}, which we refer to as LS-HRN-exact, the acronym standing for ``Leverage Scores via Hat Row Norms''. It uses Algorithm \ref{alg:count_gauss_strong_rrqr} as a preprocessing step to estimate the rank of $A$ and retrieve a set $\mathcal{K}$ of $k$ linearly independent columns. This way we avoid the computation of $A^\top A$ and instead we compute $A_{:,\mathcal{K}}^\top A_{:,\mathcal{K}}$, therefore the total complexity of the algorithm is smaller than Algorithm \ref{alg:ls_direct} when $k<d$. We also prove approximation bounds based on Theorem \ref{thm:ls_qr_vs_svd}. 
{As corroborated in the numerical experiments of Section \ref{sec:experiments}, when the singular value gap of $A$ is sufficiently large, the \rsls\ of $A_k$ are almost identical to the ones obtained by selecting a linearly independent set of columns. We prove the following theorem.}

\begin{algorithm}[htb]
	\caption{LS-HRN-exact$(A,\zeta)$}\label{alg:ls_hrn_exact}
	\begin{algorithmic}[1]
		\Require matrix $A \in \mathbb{R}^{n \times d}$, reciprocal condition number tolerance $\zeta$.
		\Ensure $\tilde\theta_i(A_k)$, {approximate leverage scores of $A_k$ where $k\leq d$ is the numerical rank of $A$. If $k=\text{rank}(A)$ then $\tilde \theta_i=\theta_i(A_k)$.}
        \State Set $\gamma_p=2,\epsilon_p=0.5,\delta_p=1/3$
        \State Compute $P_k,R_k,k\leftarrow$CountGaussSRRQR($A,\gamma_p,\epsilon_p,\delta_p,\zeta$) \Comment{$\mathcal{O}({\tt nnz}(A)+d^4)$}
	    \State Set $A_{:,\mathcal{K}}\leftarrow AP_k$ (keep only $k$ columns of $A$).
		\State Compute $B=A_{:,\mathcal{K}}^\top A_{:,\mathcal{K}}$. \Comment{$\mathcal{O}({\tt nnz_2}(A_{:,\mathcal{K}}))$}
    	\State Compute {the SVD of $B=V\Sigma^2V^\top$}. \Comment{$\mathcal{O}(k^3)$}
		\State \Return $\tilde \theta_i =  \| e_i^\top A_{:,\mathcal{K}} V\Sigma^{-1}\|_2^2\ i\in[n]$ \Comment{$\mathcal{O}({\tt nnz_2}(A_{:,\mathcal{K}}))$}
	\end{algorithmic}
\end{algorithm}

\begin{theorem}
\label{thm:ls_hrn_exact_bounds}
Given $A\in\mathbb{R}^{n\times d}$, if we use Algorithm \ref{alg:ls_hrn_exact}  to compute its numerical rank  $k$ and the leverage scores $\tilde \theta_i(A_k)$, with probability at least $(1-\delta)(1-2e^{-\alpha^2m/2})$ it holds that
\begin{align*}
|\theta_i(A_k) - \tilde \theta_i(A_k)|
<
\left(
    \sqrt{\theta_i(A_k)} 
    + 
    \sqrt{\tilde \theta_i(A_k)}
\right)
\dfrac{\sigma_{k+1}(A)}{\sigma_{k}(A)}
\xi\eta\rho
\end{align*}
where $\xi = \frac{(1+\alpha+\sqrt{k/m})}{(1-\alpha-\sqrt{k/m})}$, $\eta=\frac{(1+\epsilon)}{(1-\epsilon)}$ and $\rho=\sqrt{1+\phi^2k(d-k)}$.
\begin{proof}
{Let $A_{:,\mathcal{K}}=Q_1R_{11}$ be a QR factorization of $A_{:,\mathcal{K}}$. From Theorem \ref{thm:ls_qr_vs_svd} we have that
\begin{align*}
|\theta_i(A_k) - \theta_i(A_{:,\mathcal{K}})|
\leq
\left(\sqrt{\theta_i(A_k)} + \sqrt{\theta_i(A_{:,\mathcal{K}})}
\right)
\dfrac{\sigma_{k+1}(A)}{\sigma_{k}(R_{11})}.
\end{align*}
Moreover, from Lemma \ref{lem:bound_sigma_min_r_11}, with probability at least $(1-\delta)(1-2e^{-\alpha^2m/2})$ it holds that 
$
\sigma_{k}(R_{11}) 
> 
\dfrac{\sigma_k(A)}{\xi\eta \rho}
$.
Combining these two inequalities concludes the proof.}
\end{proof}
\end{theorem}
{We compare Algorithms \ref{alg:ls_direct}, \ref{alg:ls_spqr} and \ref{alg:ls_hrn_exact} in Table \ref{tab:exact_lsalgs_sparse}.}

\begin{table}[htb]\footnotesize
    \centering
    \caption{Comparison of leverage scores Algorithms \ref{alg:ls_direct}, \ref{alg:ls_spqr} and \ref{alg:ls_hrn_exact} for rectangular matrices. For the complexity of \ref{alg:ls_spqr} we have assumed $\mathcal{O}({\tt nnz}(A)/d)$ non zero elements per column of $A$. $\tilde \theta_i$ and $A_{:,\mathcal{K}}\in\mathbb{R}^{n\times k}$ are computed by each algorithm independently.}
    \begin{tabular}{l l l l }
        \hline
        Algorithm &
        Alg. \ref{alg:ls_direct}  &
        Alg. \ref{alg:ls_spqr} & 
        Alg. \ref{alg:ls_hrn_exact} \\\hline\hline
        Sparse & $\mathcal{O}({\tt nnz_2}(A)+d^3)$ & 
         $\mathcal{O}\left({\tt nnz_2}(A_{:,\mathcal{K}}) + {\tt nnz}(A)(1+\frac{k^2}{d})\right)$ & 
         $\mathcal{O}({\tt nnz}(A)+d^4+{\tt nnz_2}(A_{:,\mathcal{K}}))$ \\
        Dense & $\mathcal{O}(nd^2)$ & 
         $\mathcal{O}(nd+nk^2)$ & 
         $\mathcal{O}(nd+d^4+nk^2)$ \\
         \hline
         $\left|\theta_i(A_k)-\tilde\theta_i\right|$
         &
         exact
         &
         -
         &
         Theorem {\ref{thm:ls_hrn_exact_bounds}}
         \\[+0.5em]
         \hline
    \end{tabular}
    \label{tab:exact_lsalgs_sparse}
\end{table}

\subsection{Sketching based methods for dense matrices\label{sec:sketching_ls_algorithms}}

We next discuss approximate leverage scores algorithms which use sketching to reduce the complexity for tall-and-thin matrices which are sufficiently dense, by estimating \rsls\ as the squared row norms of the matrix
\begin{equation}
    A(\Pi_1 A)^\dagger \Pi_2
    \label{eq:soa_ls}
\end{equation}
where $\Pi_1\in\mathbb{R}^{r_1\times n}$ is an $\epsilon_1$-OSE  and $\Pi_2\in\mathbb{R}^{d\times r_2}$ is a $\epsilon_2$-JLT for $n$ vectors. These methods are listed as Algorithms \ref{alg:ls_approximate_drineas} and \ref{alg:ls_approximate}.

\begin{algorithm}[htb]
	\caption{\cite{drineas2012fast}}\label{alg:ls_approximate_drineas}
	\begin{algorithmic}[1]
		\Require matrix $A \in \mathbb{R}^{n \times d}$ with rank$(A)=d$.
		\Ensure $\tilde \theta_i$, approximate leverage scores of $A$.
		\State Choose two sketching matrices $\Pi_1\in\mathbb{R}^{r_1\times n},\Pi_2\in\mathbb{R}^{d\times r_2}$.
        \State Compute $\tilde A\leftarrow \Pi_1A$ \Comment{$\mathcal{O}({\tt nnz}(A))$ for a Countsketch $\Pi_1$.}
        \State Compute either $\tilde \Sigma,\tilde V$ from the SVD of $\tilde A$ or $\tilde R$ from a QR.\Comment{$\mathcal{O}(r_1d^2)$.}

	    \State Compute $X\leftarrow \tilde V \tilde \Sigma^{-1}\Pi_2$. \Comment{$\mathcal{O}(d^2\log r_2)$ for a SRHT $\Pi_2$.}
	    \State Return $\tilde\theta_i\leftarrow \|e_i^\top AX\|_2^2$. \Comment{$\mathcal{O}(\min\{{\tt nnz_2}(A),{\tt nnz}(A)r_2\})$.}
	\end{algorithmic}
\end{algorithm}

\begin{algorithm}[htb]
	\caption{\cite{clarkson2013low_jacm,nelson2013osnap}}\label{alg:ls_approximate}
	\begin{algorithmic}[1]
		\Require matrix $A \in \mathbb{R}^{n \times d}$.
		\Ensure $\tilde \theta_i$, approximate leverage scores of $A$.
		\State Use the Algorithm from {Corollary \ref{cor:cheung_complexity_tall_and_thin} on $A$} to obtain $A_{:,\mathcal{K}}$, {where $\mathcal{K}$ defines a maximal set of linearly independent columns of $A$}. \Comment{$\mathcal{O}({\tt nnz}(A)+d^\omega)$.}
        \State Compute $\tilde \theta_i(A_{:,\mathcal{K}})$ using Algorithm \ref{alg:ls_approximate_drineas} on $A_{:,\mathcal{K}}$. \Comment{$\mathcal{O}(\min\{{\tt nnz_2}(A_{:,\mathcal{K}}),{\tt nnz}(A_{:,\mathcal{K}})r_2\})$.}
	    \State Return $\tilde \theta_i(A_{:,\mathcal{K}})$. 
	\end{algorithmic}
\end{algorithm}

{The underlying idea is that, when the matrix is sufficiently dense, we can reduce the cost of computing an orthogonalizer and the squared row norms, by computing them approximately. For the orthogonalizer, instead of forming $B=A^\top A$ and {its pseudoinverse} in $\mathcal{O}({\tt nnz_2}(A)+d^3)$ operations, they compute $B=\Pi_1A$ and its SVD $B=U\Sigma V^\top$ in $\mathcal{O}(r_2d^2)$ operations. If $\Pi_1$ is a CountSketch, then this translates to a total of $\mathcal{O}({\tt nnz}(A)+d^4)$ operations, which can be smaller than $\mathcal{O}({\tt nnz_2}(A)+d^3)$ if $A$ is sufficiently dense. Similarly, for the row norms, instead of computing the squared row norms of $AQ\Lambda^{-1/2}$ in $\mathcal{O}({\tt nnz_2}(A))$ they compute the row norms of $AV\Sigma^{-1}\Pi_2$ in $\mathcal{O}({\tt nnz}(A)r_2)$, which is again faster if $A$ is sufficiently dense.}

\subsection{Sampling based methods}
The leverage scores algorithms that we described so far are based on random projection methods. Sampling based methods have also been studied in literature \cite{Cohen:2015cb,cohen2017input,Li:2013is}.
We cite \cite[Alg. 2]{Cohen:2015cb} as an example.
For further details see \cite[Lemmas 3, 7 and 10]{Cohen:2015cb}. The complexity of this algorithm is 
\begin{equation}
    \label{eq:complexity_cohen_ls}
    \mathcal{O}\left({\tt nnz}(A)\psi^{-1}+d^\omega\log^2d+d^{2+\psi}(\log^3 d + \epsilon^{-2}\log^2d)\right)
\end{equation} for some given $\psi\in(0,1]$ and $\epsilon\in(0,1)$, with probability high in $d$, outputs a matrix $\tilde A$, consisting of $\mathcal{O}(\epsilon^{-2}d\log d)$ rows of $A$ such that
for the values 
\begin{align*}
\tilde \theta_i 
= 
\left\{\begin{array}{l}
A_{i,:}^\top(\tilde A^\top \tilde A)^\dagger A_{i,:},\text{ if }A_{i,:}\in\text{range}(\tilde A^\top \tilde A)\\
1,\text{ else }
\end{array}\right.
\end{align*}
it holds that 
$\theta_i(A)\leq\tilde\theta_i\leq(1+\epsilon)\theta_i(A).$

These values are expensive to compute, but one can derive approximations $\tilde \theta_i \leq \hat\theta_i\leq d^\psi\tilde\theta_i$ in time $\mathcal{O}({\tt nnz}(A)\psi^{-1})$. This leads to a total error of $\mathcal{O}(d^\psi(1+\epsilon)).$
Since this algorithm is targeted for $\epsilon$-spectral approximations of $A$, crude estimations (i.e. large $\psi$) of these generalized leverage scores are sufficient. However, for tighter approximations, $\psi$ cannot be too large. In that case, based on \cite{Cohen:2015cb} and our analysis in Section \ref{sec:curse}, we observe that the term $\mathcal{O}({\tt nnz}(A)\psi^{-1})$ in Equation \ref{eq:complexity_cohen_ls} is in fact 
\begin{align*}\mathcal{O}\left(\min\{{\tt nnz}(A)\psi^{-1},{\tt nnz_2}(A)\}\right).\end{align*}
This means that if $A$ is very sparse, the total complexity of this algorithm is also affected by the expensive row norms computation of Lemma \ref{lem:row_norms_complexity}, just like the rest of the algorithms discussed.

\subsection{Algorithm LS-HRN-approx}  

Algorithm \ref{alg:ls_approximate} uses \cite[Theorems $2.7$ and $2.11$]{cheung2013fast} as a preprocessing step,  to output a matrix $A_{:,\mathcal{K}}\in\mathbb{R}^{n\times k}$ consisting of $k$ linearly independent columns of $A$ { and in Corollary \ref{cor:cheung_complexity_tall_and_thin} we showed that the cost for this preprocessing step is $\mathcal{O}({\tt nnz}(A)+d^\omega)$ for tall-and-thin matrices. 
We propose that
we can alternatively use Algorithm \ref{alg:count_gauss_strong_rrqr} to recover $A_{:,\mathcal{K}}$ in $\mathcal{O}({\tt nnz}(A)+d^4)$. Doing so, we are able to select $k<\text{rank}(A)$ columns.}
We describe this in Algorithm \ref{alg:ls_hrn_approx}, which is derived from Algorithm \ref{alg:ls_approximate} by substituting the 
column selection step.

\begin{algorithm}[htb]
	\caption{LS-HRN-approx}\label{alg:ls_hrn_approx}
	\begin{algorithmic}[1]
		\Require matrix $A \in \mathbb{R}^{n \times d}$, 
		reciprocal condition number tolerance $\zeta$.
		\Ensure $\tilde \theta_i(A_k)$, approximate \rsls\  of $A_k$, where $k\leq$rank$(A)$  is the numerical rank of $A$.
		\State Set $\gamma_p=2,\epsilon_p=0.5,\delta_p=1/3$
        \State Compute $P_k,R_k,k\leftarrow$CountGaussSRRQR($A,\gamma_p,\epsilon_p,\delta_p,\zeta$). \Comment{$\mathcal{O}({\tt nnz}(A)+d^4)$.}
	    \State Set $A_{:,\mathcal{K}}\leftarrow AP_k$ (keep only the best $k$).
        \State Compute $\tilde \theta_i(A_{:,\mathcal{K}})$ using Algorithm \ref{alg:ls_approximate_drineas} on $A_{:,\mathcal{K}}$. \Comment{$\mathcal{O}{\left(\min\{{\tt nnz_2}(A_{:,\mathcal{K}}),{\tt nnz}(A_{:,\mathcal{K}})r_2\}\right)}$.}
	    \State Return $\tilde \theta_i(A_{:,\mathcal{K}})$.
	\end{algorithmic}
\end{algorithm}

\begin{table}[htb]
    \centering
    \small
    \caption{Comparison of Algorithms \ref{alg:ls_approximate} and \ref{alg:ls_hrn_approx}. Here $\mathcal{K}\subset [d],|\mathcal{K}=k$, is computed by each algorithm independently, for some given inputs $k\leq d$ and $\epsilon\in(0,1)$.}
    \label{tab:ls_approximate_cost_breakwodn}
    \begin{tabular}{c c  c  c  }
    \hline
        \multirow{2}{*}{Algorithm }
        & 
        \multirow{2}{*}{Complexity }
        & 
        \multicolumn{2}{c}{$\left|\tilde \theta_i - \theta_i(A_k)\right|$} 
        \\[+0.5em]\cline{3-4}
        & 
        &  
        $k=$rank$(A)$ 
        & 
        $k<$rank$(A)$
        \\\hline
        Alg. \ref{alg:ls_approximate} 
        &
        $\mathcal{O}\left({\tt nnz}(A)+d^\omega + {\tt nnz}(A_{:,\mathcal{K}})\epsilon^{-2}\right)$ 
        &
        $\epsilon\theta_i(A_k)$
        &
        -
        \\\hline
        Alg. \ref{alg:ls_hrn_approx}
        &
        $\mathcal{O}\left({\tt nnz}(A)+d^4+{\tt nnz}(A_{:,\mathcal{K}})\epsilon^{-2}\right)$ 
        &
        $\epsilon\theta_i(A_k)$
        &
        Theorem \ref{thm:ls_hrn_approx_bounds}
        \\\hline
    \end{tabular}
\end{table}

The estimated values returned by Algorithms \ref{alg:ls_approximate} and \ref{alg:ls_hrn_approx}, which are based on Equation \ref{eq:soa_ls}, are bounded by $1\pm\mathcal \mathcal{O}(\epsilon_1+\epsilon_2)$. The values for $r_1, r_2$ depend on the OSE matrix type {and on $\epsilon_1^{-2}$ and $\epsilon_2^{-2}$ respectively}. Gaussian embeddings, SRHT and CountSketch are typical choices for $\Pi_1$ and $\Pi_2$. The choice of $\Pi_1, \Pi_2$ 
defines the complexity of such algorithms, which includes the cost of computing $\Pi_1A$ and $\Pi_2A$, a SVD (or QR) decomposition of $\Pi_1 A\in\mathbb{R}^{r_1\times d}$, and the final computation of the squared row norms $\|e_i^\top A((\Pi_1 A)^\dagger \Pi_2)\|_2^2, \forall i\in[n]$. We compare Algorithms \ref{alg:ls_approximate} and \ref{alg:ls_hrn_approx} in Table \ref{tab:ls_approximate_cost_breakwodn}.

\subsubsection{Approximation bounds for Algorithm \ref{alg:ls_hrn_approx}}

We are now in a position to complete our analysis by proving approximation bounds for Algorithm \ref{alg:ls_hrn_approx}.

\begin{theorem}
\label{thm:ls_hrn_approx_bounds}
Given $A\in\mathbb{R}^{n\times d}$, if we use Algorithm \ref{alg:ls_hrn_approx} to compute its numerical rank $k$ and the leverage scores $\tilde \theta_i(A_k)$, with probability at least $(1-\tilde \delta)(1-2e^{-\alpha^2m/2})$ it holds that
\begin{align*}
|\tilde \theta_i(A_k) - \theta_i(A_k)| < \tilde \epsilon \theta_i(A_k) + (1+\tilde\epsilon) \ \left(\sqrt{\theta_i(A_k)} 
+ 
\sqrt{\tilde \theta_i(A_k)}\right)\dfrac{\sigma_{k+1}(A)}{\sigma_{k}(A)}\xi\eta\rho
\end{align*}
where {$\xi = \frac{(1+\alpha+\sqrt{k/m})}{(1-\alpha-\sqrt{k/m})}$, $\eta=\frac{(1+\epsilon)}{(1-\epsilon)}$, $\rho=\sqrt{1+\phi^2k(d-k)}$,}
$(1-\tilde\delta) = (1-\delta_1)(1-\delta_2)(1-\delta)$,$\tilde \epsilon = \mathcal{O}(\epsilon_1+\epsilon_2)$ and $\alpha\in(0,1-\sqrt{k/m})$.
\begin{proof}
We write 
\begin{align*}
|\tilde \theta_i(A_k) - \theta_i(A_k)|&=|\tilde \theta_i(A_k)-\theta_i(A_k) +\theta_i(A_{:,\mathcal{K}}) - \theta_i(A_{:,\mathcal{K}})| \\
&\leq |\tilde \theta_i(A_k)- \theta_i(A_{:,\mathcal{K}})| + |\theta_i(A_{:,\mathcal{K}}) -\theta_i(A_k)|.
\end{align*}
We bound each term separately.
For the leftmost term, we have that 
\begin{align*}|\tilde \theta_i(A_k)- \theta_i(A_{:,\mathcal{K}})|\leq \tilde \epsilon \theta_i(A_{:,\mathcal{K}})\end{align*}
from \cite[Theorem 2]{drineas2012fast}, for some $\tilde \epsilon = \mathcal{O}(\epsilon_1+\epsilon_2)$. Moreover, from the triangle inequality we have that $\theta_i(A_{:,\mathcal{K}})\leq \theta_i(A_k)+|\theta_i(A_k)-\theta_i(A_{:,\mathcal{K}})|$ which leads to
\begin{align*}|\tilde \theta_i(A_k)- \theta_i(A_{:,\mathcal{K}})|\leq \tilde \epsilon \theta_i(A_k)+\tilde\epsilon|\theta_i(A_k)-\theta_i(A_{:,\mathcal{K}})|.
\end{align*}
From Theorem \ref{thm:ls_hrn_exact_bounds} we have an immediate bound for $|\theta_i(A_{:,\mathcal{K}}) -\theta_i(A_k)|$, finally leading to
\begin{align*}
|\tilde \theta_i(A_k) - \theta_i(A_k)| 
< 
\tilde \epsilon \theta_i(A_k) 
+ 
(1+\tilde\epsilon) \left(\sqrt{\theta_i(A_k)} 
+ 
\sqrt{\tilde \theta_i(A_k)}\right)
\dfrac{\sigma_{k+1}(A)}{\sigma_{k}(A)}\xi\eta\rho.
\end{align*}
The total success probability equals to the product of the success probabilities of the subspace embeddings.
\end{proof}
\end{theorem}

\subsection{Overall evaluation\label{sec:ls_overall}}
It is now possible to rank all the algorithms from Tables \ref{tab:exact_lsalgs_sparse} and \ref{tab:ls_approximate_cost_breakwodn} based on their complexities ($a\succ b$ means that $a$ has smaller complexity than $b$).
\begin{description}
\item[Dense matrices]
\begin{align*}
\text{Alg. \ref{alg:ls_hrn_approx}} \succ \text{\cite[Alg. 2]{Cohen:2015cb}}  \succ \text{Alg. \ref{alg:ls_approximate} } \approx \text{Alg. \ref{alg:ls_hrn_exact}} \succ \text{Alg. \ref{alg:ls_direct} }
\end{align*}
\item[Sparse matrices] \begin{align*}
\text{Alg. \ref{alg:ls_hrn_exact}} = \text{Alg. \ref{alg:ls_hrn_approx} } \approx \text{\cite[Alg. 2]{Cohen:2015cb}} \succ \text{Alg. \ref{alg:ls_approximate} } \succ \text{Alg. \ref{alg:ls_direct} }
\end{align*}
\end{description}

Observe that, if $k$ is very small, i.e. $k\ll \sqrt{d}$, then in terms of complexity Algorithm \ref{alg:ls_spqr} becomes a competitive candidate as it will only process a few columns of $A$. However, due to the greedy column selection strategy, the estimated values might be arbitrarily far from the leverage scores of $A_k$.

Our assumption throughout this work was that the underlying matrices are tall-and-thin. 
If this is not the case, it is worth noting that efficient algorithms and approximation bounds for can be found in the literature  \cite[Alg. 5 and 6]{drineas2012fast}, \cite[Alg. 2 and 3]{gittens2013revisiting}, \cite{drineas2018structural}. 
Such algorithms approximate the dominant-$k$ subspace using techniques related to power iterations and therefore they avoid the cubic costs on $d$ required to build an orthogonalizer for $A$. However, this introduces multiplicative factors to the $\mathcal{O}({\tt nnz}(A))$ terms, which becomes inefficient for very tall matrices.

\section{Least squares preconditioning}
As a byproduct of our analysis in the previous sections, we show that the proposed methods can be also used to derive effective preconditioners for overdetermined least squares problems.

\subsection{Preconditioning and rank estimation with Gaussian embeddings}
Random projections have been successfully used to construct preconditioners for least squares problems, and their combination with iterative methods has successfully outperformed standard high performance solvers \cite{avron2010blendenpik,Drineas2010,meng2014lsrn,RokhlinTygert.08}. {To the best of our knowledge, the first such work is \cite{RokhlinTygert.08} where the authors also discuss novel ways to choose starting vectors for the iterative solver.}

We build upon the results of Meng et al \cite{meng2014lsrn}. The authors construct the matrix product $GA$, $A\in\mathbb{R}^{n\times d}$, and $G\in\mathbb{R}^{m\times n}$ is a Gaussian matrix, with $m=\lceil \gamma d\rceil$ and $\gamma\in(1,3]$. The following results are proved.

\begin{theorem}
	\label{thm:cond} \cite[Thms. 4.1 and 4.4]{meng2014lsrn}
	Consider the matrix product $GA$, $G\in\mathbb{R}^{m\times n},A\in\mathbb{R}^{n\times d}$ where $G$ is a Gaussian matrix, $m>d$ and rank$(A)=k\leq d$, and the economy SVD of $GA=\tilde U \tilde \Sigma \tilde V^\top$, where {$\tilde U\in\mathbb{R}^{n\times k}, \tilde  \Sigma\in\mathbb{R}^{k\times k}, \tilde  V\in\mathbb{R}^{d\times k}$}. If we set $N=\tilde V\tilde \Sigma ^{-1}$, $N\in \mathbb{R}^{d\times k}$, then for any $\alpha \in (0,1-\sqrt{k/m})$ the following inequality holds:
	\begin{equation}
	    \mathbb{P}
	    \left[
	        \kappa(AN)
	        \leq 
	        \dfrac{(1+\alpha+\sqrt{k/m})}{(1-\alpha-\sqrt{k/m})}
        \right]
        \geq
        1-2e^{-\alpha^2m/2}.
	\label{eq:cond_NA}
	\end{equation}
{Moreover, given $b\in\mathbb{R}^{n}$, if we denote by $\hat y$ the minimum length solution of $\min_y\|ANy-b\|$, then for the vector $\hat x=N \hat y$ it holds that $\hat x=A^\dagger b$ almost surely.}
\end{theorem}

It is also shown in \cite{meng2014lsrn} that $GA$ can be used to efficiently identify the numerical rank of $A$. Specifically, the following theorem is proven, which intuitively states that if the singular values $\sigma_k(A)$ and $\sigma_{k+1}(A)$ are well separated outside the distortion that is introduced by the Gaussian embedding, then the numerical rank of $A$ will be equal to the numerical rank of $GA$.

\begin{theorem}
	\label{thm:rankest} 
	\cite[Thm. 4.6]{meng2014lsrn}
	Assume $G$ and $A$ to be as in Theorem \ref{thm:cond} and that there exists a small constant $c>0$ such that $\sigma_1(A)\geq \sigma_k(A)\gg c\sigma_1(A) \gg \sigma_{k+1}(A)$. If $G$ satisfies the subspace embedding property, then there exist constants $q_1,q_2$ such that $q_1\|G x\|\leq \| x\|\leq q_2\|G x\|$ for all $ x\in \text{range}(A)$. If it holds also that $\sigma_{k+1}(A)\leq c\sigma_1(A)q_1/q_2 $ and $\sigma_k(A)>c(1+q_2/q_1)\sigma_1(A)$, then $\tilde k = k$ where $\tilde k$ is the numerical rank of $GA$, i.e. the rank of $GA$ if we truncate all the singular values which are smaller than $c\sigma_1(GA)$.
\end{theorem}

\subsection{Preconditioning with CountGauss}
{Inspired by the results in \cite{clarkson2013low_jacm,dahiya2018empirical,meng2014lsrn,woodruff2014sketching}, we will next use Algorithm \ref{alg:count_gauss_sketch} to build least squares preconditioners. We first prove a useful lemma.
\begin{lemma}
\label{lem:spectrum_of_AN}
Given $A\in\mathbb{R}^{n\times d}$ with rank$(A)=k\leq d$ and its economy SVD $A=U\Sigma V^\top$, let $\tilde G=GS$ where $G\in\mathbb{R}^{m\times r}$ is a Gaussian matrix and $S\in\mathbb{R}^{r\times n}$ is an $(\epsilon,\delta)$-OSE with $\epsilon,\delta\in(0,1)$. Let the economy SVD of $\tilde GA=\tilde U \tilde \Sigma \tilde V^\top$, where $\tilde U\in\mathbb{R}^{n\times k}, \tilde  \Sigma\in\mathbb{R}^{k\times k}, \tilde V\in\mathbb{R}^{d\times k}$, and let $N=\tilde V\tilde \Sigma ^{-1}$. The spectrum of $AN$ is the same as the spectrum of $(GSU)^\dagger$ independent of the spectrum of $A$.
\begin{proof}
    The result follows from the proof of \cite[Lemma 4.2]{meng2014lsrn} simply by replacing $G_1=GU$ with $G_1=GSU$ in the original Lemma.
\end{proof}
\end{lemma}
}

\begin{theorem}
\label{thm:cond_gsa}
    Given $A\in\mathbb{R}^{n\times d}$ with rank$(A)=k\leq d$, let $\tilde G=GS$ where $G\in\mathbb{R}^{m\times r}$ is a Gaussian matrix and $S\in\mathbb{R}^{r\times n}$ is an $(\epsilon,\delta)$-OSE with $\epsilon,\delta\in(0,1)$. {Let the economy SVD of $\tilde GA=\tilde U \tilde \Sigma \tilde V^\top$, where $\tilde U\in\mathbb{R}^{n\times k}, \tilde  \Sigma\in\mathbb{R}^{k\times k}, \tilde V\in\mathbb{R}^{d\times k}$, and let} $N=\tilde V\tilde \Sigma ^{-1}$. The following  inequality holds 
    \begin{equation}
        \mathbb{P}\left[
            \kappa(AN)
            \leq \xi\eta
        \right] 
        \geq (1-2e^{-\alpha^2m/2})(1-\delta)
    \end{equation}
    for any $\alpha\in(0,1-\sqrt{k/m})$, where $\xi=\frac{(1+\alpha+\sqrt{k/m})}{(1-\alpha-\sqrt{k/m})}$, $\eta=\frac{1+\epsilon}{1-\epsilon}$.
    {Moreover, given a vector $b\in\mathbb{R}^{n}$, if we denote by $\hat y$ the minimum length solution of $\min_y\|ANy-b\|$, then for the vector $\hat x=N \hat y$ it holds that $\hat x=A^\dagger b$ almost surely.}
    {
    \begin{proof}
    From Lemmas \ref{lem:spectrum_gsu} and \ref{lem:spectrum_of_AN} we have that
    \begin{align*}\kappa(AN)=\kappa((GSU)^\dagger)=\frac{\sigma_1(GSU)}{\sigma_k(GSU)}<\frac{((1+\alpha)\sqrt{m}+\sqrt{k})(1+\epsilon)}{((1-\alpha)\sqrt{m}-\sqrt{k})(1-\epsilon)}.\end{align*}
    For the solution consistency, observe that 
    $\text{range}(N)=\text{range}(\tilde V\tilde \Sigma^{-1})=\text{range}((GSA)^\top)=\text{range}(V\Sigma(GSU)^\top)=\text{range}(V)=\text{range}(A^\top)$, where we have used the fact that $GSU$ has full column rank; cf. \cite[\S 3.1]{dahiya2018empirical}, \cite[Thm. 4.1]{meng2014lsrn}, \cite{woodruff2014sketching}.
    \end{proof}}
\end{theorem}

{Algorithm \ref{alg:count_gauss_precondition} describes a preconditioning strategy based on CountGauss transforms which satisfies Theorem \ref{thm:cond_gsa}.}

\begin{algorithm}[htb]
\caption{CountGaussPrecondition$(A,\gamma,\delta,\epsilon,\zeta,\phi)$: Right preconditioning of very tall sparse matrices with CountGauss embeddings and SVD.}
\label{alg:count_gauss_precondition}
\begin{algorithmic}[1]
	\Require Matrix $A\in \mathbb{R}^{n\times d}$ with $n\gg \mathcal{O}(d^2)$, ${\tt nnz}(A)>\mathcal{O}(d^3)$, oversampling factor $\gamma \in (1,3]$, CountSketch OSE parameters $\epsilon,\delta\in(0,1)$, tolerance $\zeta$ for ${\tt rcond}(A)$.
	\Ensure preconditioner $N \in \mathbb{R}^{d\times d}$.
	\State {
    Compute $\tilde A\leftarrow $CountGaussSketch$(A, \gamma, \delta, \epsilon)$. \Comment{$\mathcal{O}({\tt nnz}(A)+\gamma d^4/(\epsilon^2\delta))$}.}
	\State Compute the SVD of $\tilde A=U\Sigma V^\top$. \Comment{$\mathcal{O}(\gamma d^3)$.}
	\State Set $k=\arg\max_{j\in[d]}\left\{ \sigma_j(A) \middle\vert \frac{\sigma_{j}(A)}{\sigma_1(A)}\leq \zeta\right\}.$
	\State Set $N\leftarrow V_{:,1:k}\Sigma_{1:k,1:k}^{-1}$.
	\State \Return $N, k$.
\end{algorithmic}
\end{algorithm}

\section{Implementations and numerical experiments\label{sec:experiments}}
We implemented the discussed algorithms in Python, using SciPy with OpenBLAS and OpenMP as a backend. We also deployed our own kernels to accommodate parallel sparse BLAS and related routines in C++ with OpenMP along with their Python wrappers. Algorithm \ref{alg:count_gauss_strong_rrqr} was implemented using the BLAS3 version of the classical pivoted QR factorization \cite{businger1965linear}, described in \cite{quintana1998blas} (LAPACK routine {\tt xGEQP3}). The algorithm is 
{available in Python via SciPy. Even though it is not as robust as Algorithm \ref{alg:srrqr} in terms of computing a RRQR factorization, it works well in practice.}

\subsection{Datasets\label{sec:datasets}}
\begin{table}[htb]
    \small
    \centering
    \caption{Rectangular matrices used in the experiments. (T) denotes that the dimensions correspond to the transpose of the original matrix. The numerical rank and the 2-norm condition number are also tabulated.}
\begin{tabular}{rl|rrrrrrr}
id & name                  & $n$          & $d$          & num. rank       & {\tt nnz}       &  $\kappa_2$ \\\hline\hline
1 & timg80                & $79,302,017$ & $1,024$      & $1,024$         & $1,586,040,340$ & - \\
2 & kl02 (T)              & $36,699$     & $71$         & $64$            & $212,536$       & $1.40e16$ \\  
3 & fixed\_svd\_1e7       & $50,000$     & $60$         & $30$            & $3,000,000$     & $1e7$ \\
4 & fixed\_svd\_2.5e4       & $50,000$     & $60$         & $60$            & $3,000,000$     & $2.5e4$ \\  
\end{tabular} 
    \label{tab:matrices}
\end{table}

We evaluate our algorithms
using sparse synthetic and
``real world'' matrices listed in Table \ref{tab:matrices}. 
The timg80 matrix was constructed based on the Tiny Images dataset \cite{torralba200880}, which consists of 80 million images of size $32\times 32$ pixels.  
Following \cite{meng2014lsrn}, we convert the images in greyscale, compute the two dimensional Discrete Cosine Transform and keep only the largest $20$ elements in magnitude. We then flatten the two-dimensional arrays to vectors and stack them vertically in arbitrary order, obtaining a sparse matrix of size $79,302,017\times 1,024$ and approximately $1.6$ billion non-zero elements. 
The matrix fixed\_svd\_1e7 was constructed such that the first fifteen singular values are equal to $1$, the following fifteen equal to $10^{-6}$ and the last thirty equal to $10^{-7}$. Similarly, the matrix fixed\_svd\_2.5e4 has the first fifteen singular values are equal to $1$, the following fifteen equal to $10^{-3}$ and the last $30$ equal to $4\times 10^{-5}$. We use these matrices as well as kl02 from SuiteSparse Matrix Collection\footnote{\href{https://sparse.tamu.edu/}{https://sparse.tamu.edu/}} to test our algorithms on numerically rank deficient input.

\subsection{Rank estimation, singular values and column subset selection}
\begin{figure}[htb]
    \centering
    \includegraphics[width=0.49\textwidth]{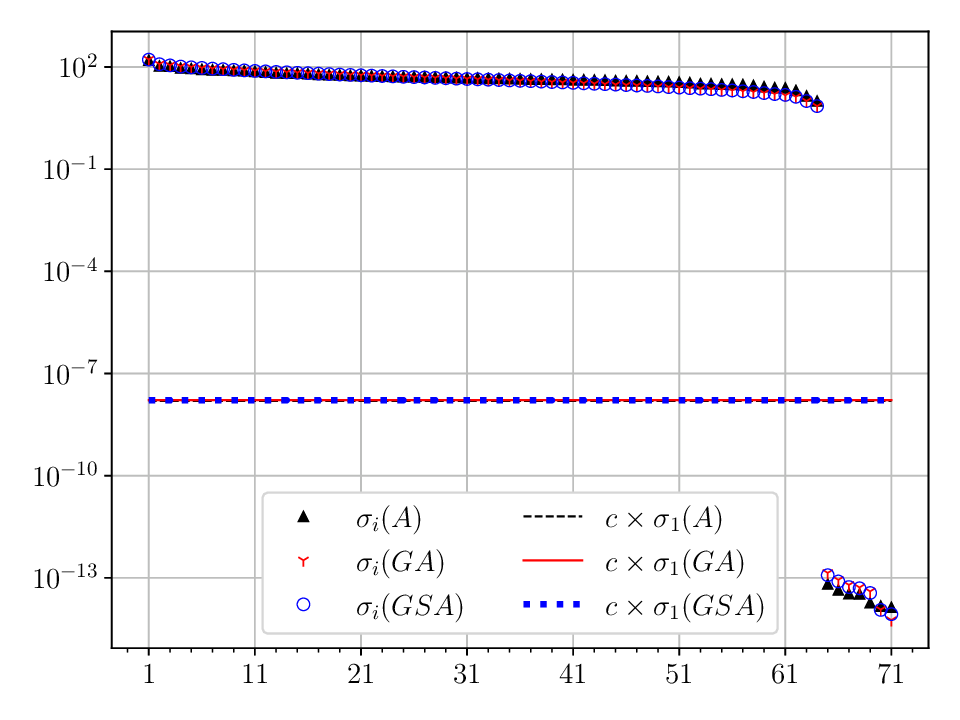}
    \includegraphics[width=0.49\textwidth]{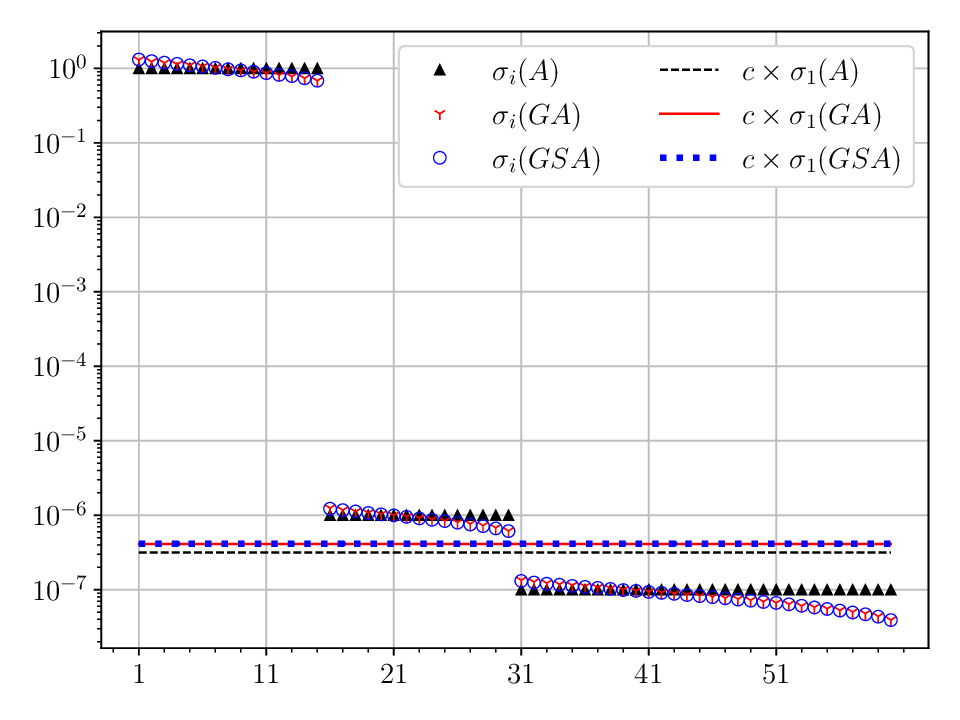}
    \caption{Singular values of $A$, $GA$ and $GSA$ for the matrix kl02 (left) and a random matrix of size $50,000\times 60$, with the first $15$ singular values equal to $1$, the next $15$ equal to $10^{-6}$ and the last $30$ equal to $10^{-7}$ (right). The solid lines are respective probabilistic upper bounds. $G$ is a Gaussian embedding with $m=\lceil\gamma d\rceil$ rows, $\gamma=2$, and $S$ is a CountSketch where we set $r=5(d^2+d)$ rows. We plot the average over $20$ runs.}
    \label{fig:svd_kl02}
\end{figure}

In the next experiment we use the matrix kl02 which is numerically rank deficient. It has $71$ columns but its numerical rank is $64$. Figure \ref{fig:svd_kl02} depicts the singular values of $A$, $GA$ and $GSA$. Here $G$ is a Gaussian embedding with $m=2d$ rows and $S$ is a CountSketch with $r=5(d^2+d)$ rows, which is approximately the number of required such that $S$ will be an $(\epsilon,\delta)$-OSE with $\epsilon=0.5$ and $\delta=1/3$. For the singular values of $GA$ and $GSA$ we plot the average over $50$ independent runs. The solid lines correspond to the upper estimation bounds based on Theorems \ref{thm:cond} and \ref{thm:cond_gsa}, where we set  $\alpha$ accordingly such that for the theorems' failure probability it holds $e^{-\alpha^2m/2}\approx 0.01$. We also set a ``cutoff'' threshold $c=10^{-10}$ for kl02 and $c=10^{-6.5}$ for the other matrix, and plot the ``cutoff'' line for the singular values $c\times \sigma_1(A)$, $c\times \sigma_1(GA)$ and $c\times \sigma_1(GSA)$. The singular values of $A$ are well approximated by both $GA$ and $GSA$ and in both cases the numerical rank is returned correctly for both matrices, which are $64$ and $30$ respectively. These results indicate that, in practice, one could use fewer rows for $S$ but there is no theoretical proof to { support this (similar observations were made in \cite{dahiya2018empirical})}. 

\begin{figure}[htb]
    \centering
    \includegraphics[width=0.49\textwidth]{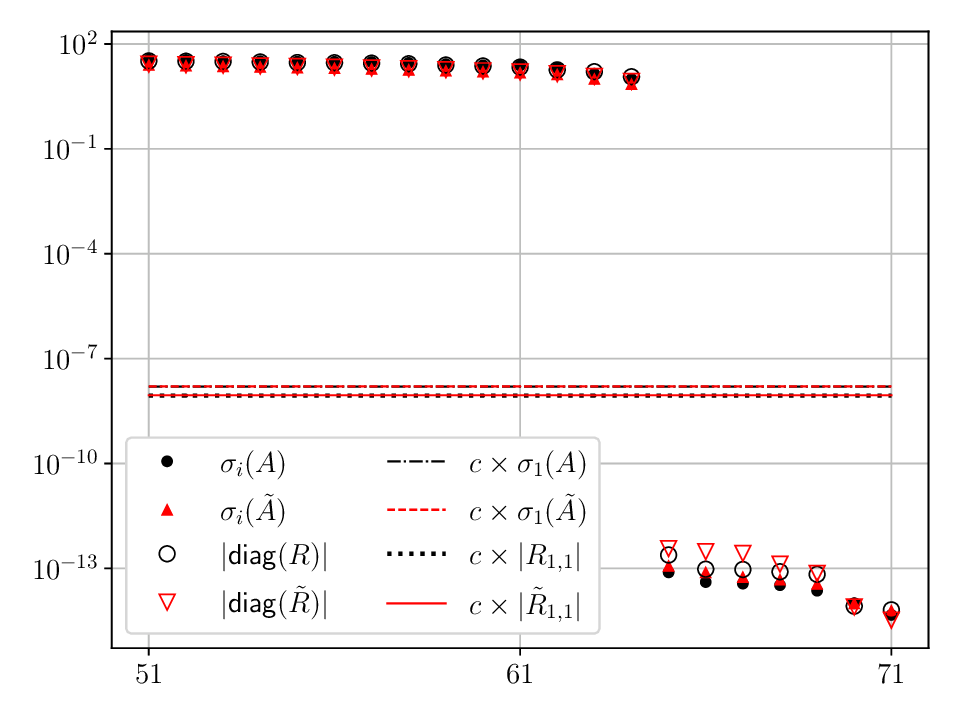}
    \includegraphics[width=0.49\textwidth]{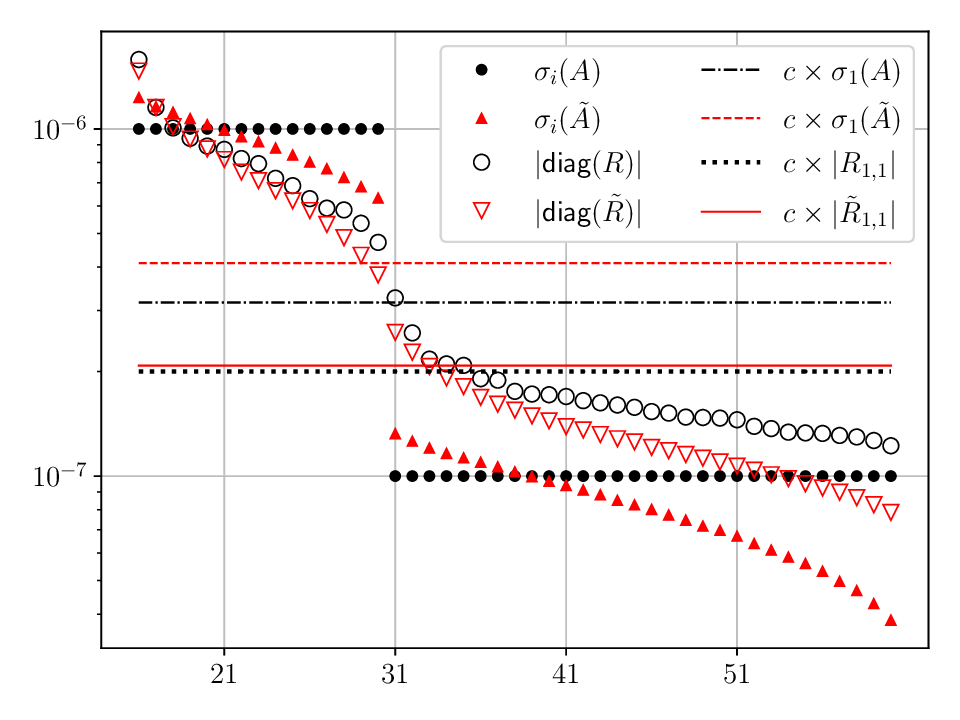}
    \caption{Running pivoted QR on $A$ and $\tilde A=GSA$ for the matrices kl02 (left) and fixed\_svd\_1e7 (right), where $G$ is a Gaussian embedding with $m=\lceil\gamma d\rceil$ rows, $\gamma=2$, and $S$ is a CountSketch where we set $r=5(d^2+d)$ rows, which is roughly the number of rows required for an  $(\epsilon,\delta)$-OSE with $\epsilon=0.5$ and $\delta=1/3$. }
    \label{fig:rrqr_kl02}
\end{figure}

We next test the ability of Algorithm \ref{alg:count_gauss_strong_rrqr} to identify the numerical rank. For the pivoted QR method we used the driver {\tt DGEQP3} from LAPACK. Although there exist more advanced algorithms, this algorithm works well in practice. The implementation does not support early stopping but the true rank of $A$ can be inferred from the diagonal of $R$. For the same matrices as those of Figure \ref{fig:svd_kl02}, we run the {\tt DGEQP3} algorithm on $A$ and on $\tilde A = GSA$, where $GSA$ is as before, obtaining the factorizations $AP=QR$ and $\tilde A\tilde P = \tilde Q \tilde R$. In Figure \ref{fig:rrqr_kl02} we plot the smallest singular values of the matrices $A$ and $\tilde A$  as well as the absolute values of the diagonal of $R$ and $\tilde R$. 
The cutoff value $c$ was set to $c=10^{-10}$ for kl02 and $c=10^{-6.5}$ for fixed\_svd\_1e7. For kl02, the singular values are well separated, and the algorithm correctly estimates the rank to be $64$, by examining if $|\tilde R_{k,k}|<c\times|\tilde R_{1,1}|$. This check fails for fixed\_svd\_1e7 which has a smaller singular value gap, suggesting that for such cases it is more efficient to use SVD to estimate the rank, and only use pivoted QR as a subsequent step to obtain the permutation corresponding to the $k$ linearly independent columns.

\subsection{Preconditioning\label{sec:experiments_preconditioning}}

\begin{figure}[htb]
    \centering
    \includegraphics[width=0.95\textwidth]{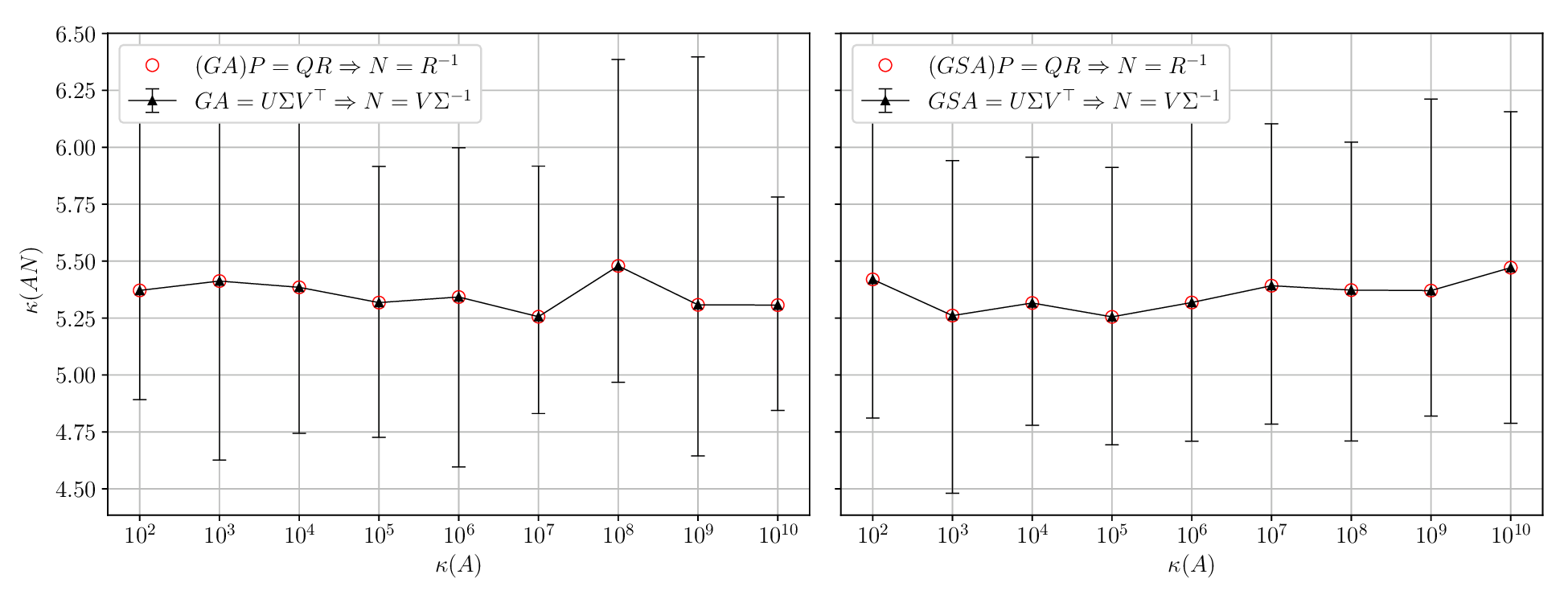}
    \caption{Comparing $\kappa(A)$ with $\kappa(AN)$ where $N$ is a preconditioner constructed with Algorithm \ref{alg:count_gauss_strong_rrqr}. The legend denotes the method used to construct $N$. We plot the mean value of $\kappa(AN)$, while the error bars present the maximum and minimum value of $\kappa(AN)$ over $20$ independent runs for each matrix.}
    \label{fig:preconditioning}
\end{figure}

We next evaluate the effect of the preconditioning suggested by Theorem \ref{thm:cond_gsa}. We construct a preconditioner $N$ using Algorithm \ref{alg:count_gauss_strong_rrqr} for dense matrices of size $50,000\times 60$ with fixed singular value distributions, equal to ${\tt linspace}(1, r, 60)$ with $r=\{10^{-2}, 10^{-3}, 10^{-4}, ..., 10^{-10}\}$. More specifically, we construct the matrices $GA$ and $GSA$ where $G$ is a Gaussian embedding with $m=\lceil\gamma d\rceil$ rows, $\gamma=2$, and $S$ is a CountSketch where we set $r=5(d^2+d)$ rows. For these matrices, we construct $N$ in two different ways, one by using SVD and one by using pivoted QR.
In Figure \ref{fig:preconditioning} we plot the results over $20$ independent runs for each matrix. We compare the condition number of $AN$ (y-axis) with the condition number of the original matrix $A$ (x-axis). We plot the mean value while the error bars indicate the maximum and minimum value for $\kappa(AN)$ over the $20$ different runs. From the plot it is evident that the condition number of $AN$ is identical if we either use QR or SVD to construct the preconditioner. Moreover, $GSA$ has almost the same strong preconditioning results with $GA$, and both cases are independent of $\kappa(A)$.

\subsection{Leverage scores}
\begin{figure}[htb]
    \centering
    \includegraphics[width=0.49\textwidth]{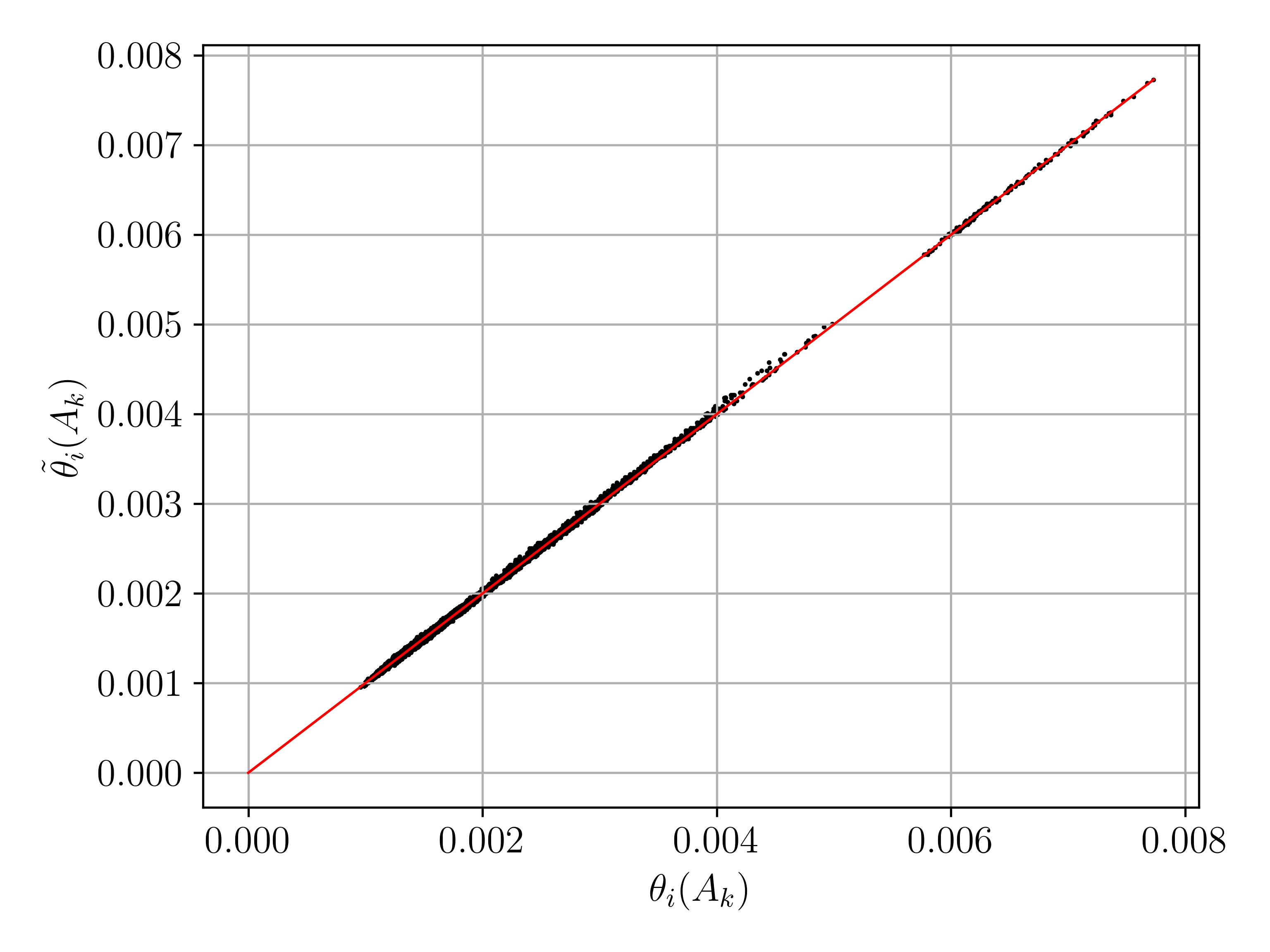}
    \includegraphics[width=0.49\textwidth]{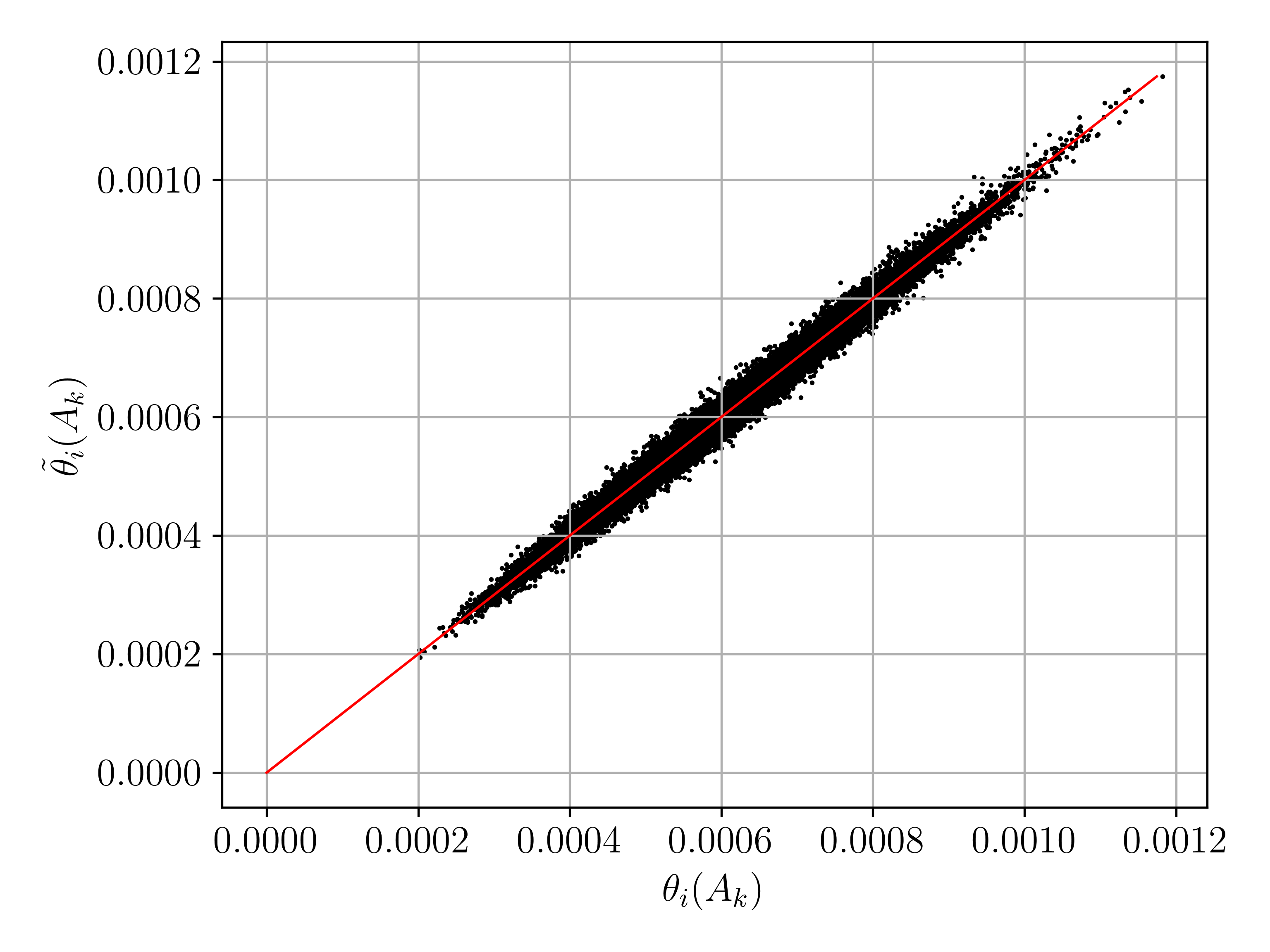}
    \caption{Comparing the leverage scores distributions $\theta_i(A_k)$ and $\tilde\theta_i(A_k)$ returned by Algorithms \ref{alg:ls_direct} and \ref{alg:ls_hrn_exact} respectively, for the matrices kl02 (left) and fixed\_svd\_2.5e4 (right).}
    \label{fig:ls_exact_comparison}
\end{figure}

We next evaluate the algorithms from Section \ref{sec:ls_algorithms}. Figure \ref{fig:ls_exact_comparison} shows a scatter plot of the leverage scores returned by  Algorithms \ref{alg:ls_hrn_exact} and \ref{alg:ls_direct} for two matrices with different properties. The left panel shows the results for the leverage scores of the kl02 matrix, which has a large singular value gap and its numerical rank is well defined. In that case the two algorithms return almost identical values. The right panel shows the results for the matrix fixed\_svd\_2.5e4. Even if the matrix is full rank, we deliberately set a large threshold equal to $2\times 10^{-4}$. In this case, the numerical rank is $30$. Algorithm \ref{alg:ls_hrn_exact} successfully determines the numerical rank equal to $30$ and it returns the leverage scores of the rows after selecting a set of linearly independent columns of $A$. Algorithm \ref{alg:ls_direct}, on the other hand, returns the leverage scores of $A_k$. For this matrix, there is a noteworthy difference between the two leverage scores distributions, which is expected since the singular value gap of this matrix is small, and therefore Theorem \ref{thm:ls_hrn_exact_bounds} suggests that there could be substantial deviations. It depends on the application whether the columns that have a small contribution in the column space should be considered or not. 

\begin{figure}[htb]
    \centering
    \includegraphics[width=\textwidth]{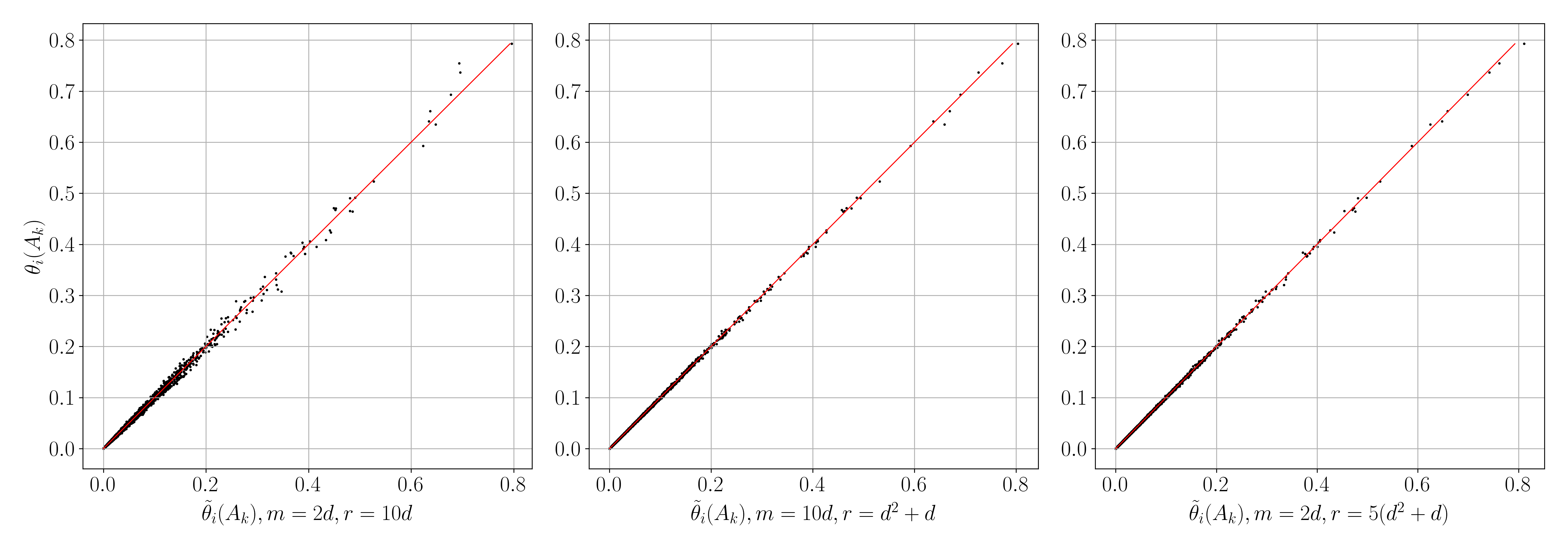}
    \caption{Comparing the leverage scores $\theta_i(A_k)$ and $\tilde \theta_i(A_k)$ returned by Algorithms \ref{alg:ls_direct} and \ref{alg:ls_approximate_drineas} respectively, for the timg80 dataset. For Algorithm \ref{alg:ls_approximate_drineas} we used $\Pi_1=GS$ where $G$ is a $m\times r$ Gaussian embedding and $S$ is a $r\times n$ CountSketch. The values for $m$ and $r$ are specified in the label of the $x$-axis in each subplot.}
    \label{fig:ls_approx_comparison}
\end{figure}

{Having examined the robustness of the algorithms to estimate leverage scores for approximately rank deficient input, we next compare exact and approximate algorithms, in particular Algorithms \ref{alg:ls_direct} and \ref{alg:ls_approximate_drineas}. In this experiment we use the timg80 dataset which happens to be full rank. For Algorithm \ref{alg:ls_approximate_drineas} we used $\Pi_1=GS$ where $G$ is a $m\times r$ Gaussian embedding and $S$ is a $r\times n$ CountSketch while we did not apply the ``$\Pi_2$'' transform, since this matrix has only $d=1024$ columns. The values for $m$ and $r$ are specified in the label of the $x$-axis in each subplot. Results are illustrated in Figure \ref{fig:ls_approx_comparison}. We observe that even by using as few as $m=2d$ rows for $G$ and $r=10d$ rows for $S$, this highly-incoherent leverage scores distribution is well approximated by Algorithm \ref{alg:ls_approximate_drineas}.}

\subsection{Performance evaluation\label{sec:experiments_performance}}
We first examine the performance of the sketching operation varying the size of the sketching matrices.
We use the timg80 dataset for our comparisons, the size of which matches the requirements of Algorithm \ref{alg:ls_hrn_exact}. These experiments were executed on a machine with Power8\footnote{\scriptsize Power8 is a trademark or registered trademark of International Business Machines Corp., registered in many jurisdictions worldwide. Other product and service names might be trademarks of IBM or other companies.}
processors, with 512GB of RAM and two CPU sockets with 10 physical cores per socket. We built our own CSR-based kernels in C++ using OpenMP for multithreading. We use $40$ OpenMP threads, equal to $2$ hyperthreads per core, which we measured to be the optimal choice for our setup.

\begin{figure}[htb]
    \centering
    \includegraphics[width=0.49\textwidth]{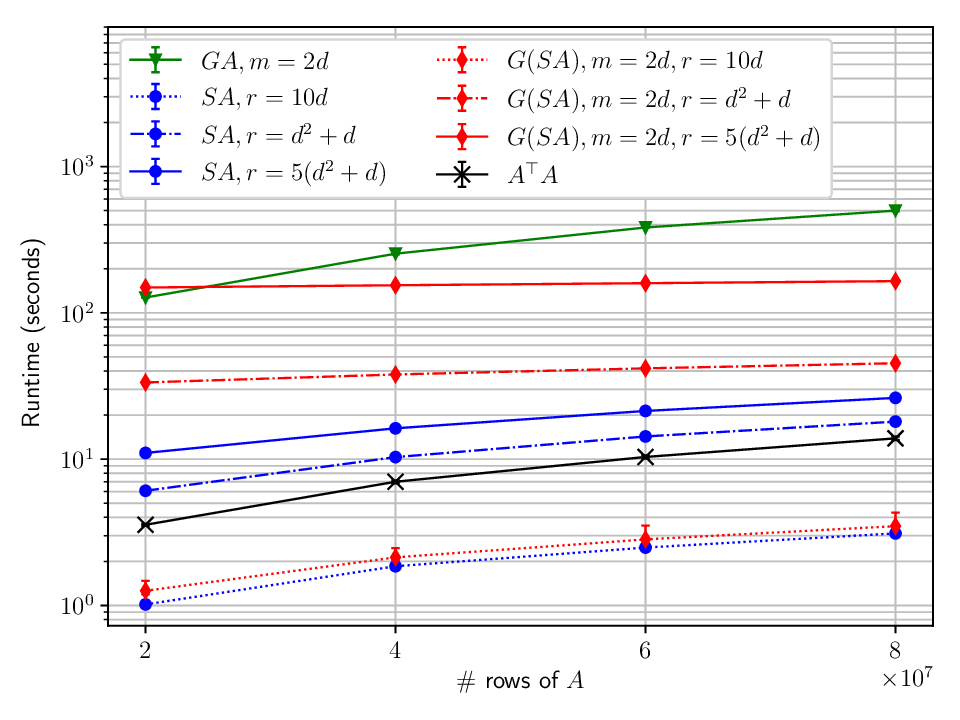}
    \includegraphics[width=0.49\textwidth]{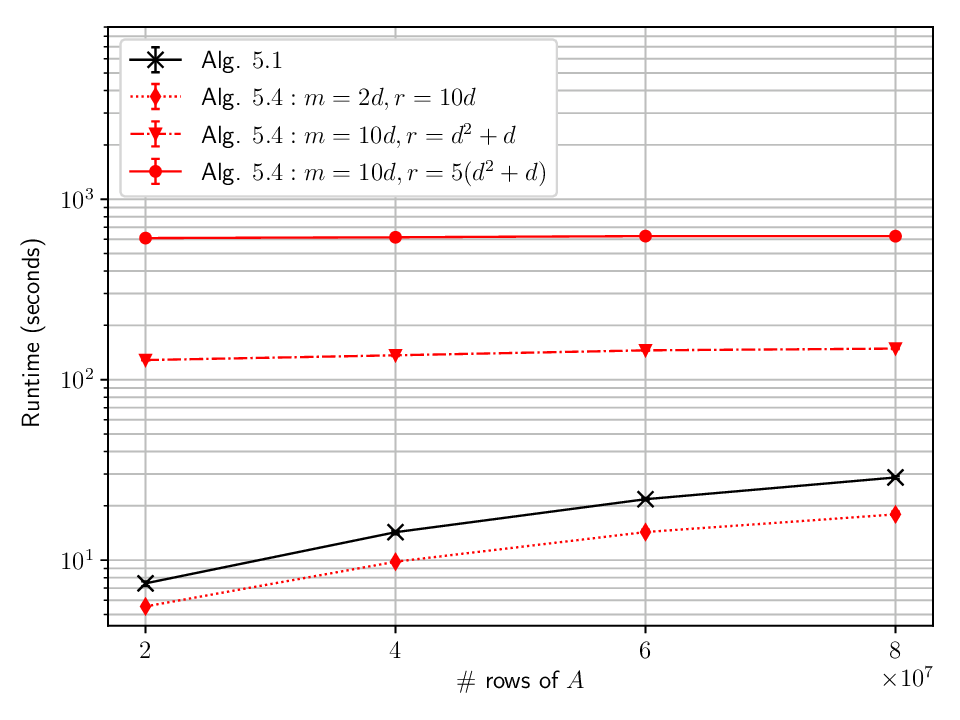}
    \caption{
    {Performance measurements for the timg80 matrix. The $x$-axis denotes how many rows of the matrix were used. For both figures $m$ is the number of rows of the Gaussian embedding $G$ and $r$ is the number of rows of the CountSketch $S$ (and the columns of $G$ respectively). The max, min and average runtimes are plotted over $5$ runs using 40 OpenMP threads. \textbf{Left}: Comparing the runtime required to compute the products $GA$, $SA$ and $GSA$. \textbf{Right}: Comparing the runtime of Algorithms \ref{alg:ls_direct} and \ref{alg:ls_approximate_drineas} to compute the leverage scores. For Algorithm \ref{alg:ls_approximate_drineas} we used $\Pi_1=GS$ and we did not apply the transform $\Pi_2$.}}
    \label{fig:ga_gsa_comparison}
\end{figure}

In Figure \ref{fig:ga_gsa_comparison} (left) we plot the runtimes for three different sketch types. In all cases, $G$ is a Gaussian embedding with $m$ rows and $S$ is a CoutSketch with $r$ rows. We display the results for different values of $m$ and $r$. For $GSA$ we first construct the entire $S$ as a sparse data structure of size $\mathcal{O}(n)$ and then we perform the computation in batches, i.e. $G_1(S_1A) + G_2(S_2A) + ... +G_b(S_bA)$, where $b$ is the number of batches, keeping in memory only one batch of $G$ at a time. $S_iA$ is computed as a first step with a sparse CSR-based matrix multiplication and the result is redundantly stored as a dense matrix, say $B_i$. Then $G_iB_i$ is computed with standard dense matrix multiplication. For larger values of $r$, the products $G_iB_i$ dominate the total runtime of the entire product and therefore there is a pseudo-constant scaling as the problem size grows. The cost of computing $GA$ grows linearly with the problem size and is orders of magnitude larger than the cost of computing $A^\top A$ which is computed based on Lemma \ref{lem:gram_complexity}. $SA$ and $GSA$ are the fastest of all the cases examined for small values of $r$. These observations can be inspiring for the design of novel HPC least squares solvers.

{On the right side of Figure \ref{fig:ga_gsa_comparison} we illustrate the runtime of Algorithms \ref{alg:ls_direct} and  \ref{alg:ls_approximate_drineas} for the same choices of $m$ and $r$ as in Figure \ref{fig:ls_approx_comparison}. For Algorithm \ref{alg:ls_approximate_drineas} we did not apply the transform $\Pi_2$ because the matrix has only $d=1024$ columns and therefore it did not provide any significant improvement. Algorithm \ref{alg:ls_direct} is simple and very fast in practice, as expected based on Lemmas \ref{lem:row_norms_complexity} and \ref{lem:gram_complexity}. On the other hand, Algorithm \ref{alg:ls_approximate_drineas} is orders of magnitude slower if we choose $m$ and $r$ such that the OSE requirements are satisfied. If, however, we select smaller values for $m$ and $r$, e.g. $m=2d$ and $r=10d$, then the Algorithm \ref{alg:ls_approximate_drineas} is faster than Algorithm \ref{alg:ls_direct} and, as illustrated in Figure \ref{fig:ls_approx_comparison}, it returns adequate approximations.}

Finally, we highlight the practical importance of using both transforms $G$ and $S$. One could argue that, since $G$ is expensive to construct and to multiply with another matrix, it might be more efficient to only use $S$. However, $S$ requires many rows, and moreover the product $SA$ will be dense and unstructured. For example, for the timg80 matrix, choosing $S$ to be a CountsSketch with $ 5d^2+d$ rows then the product $SA$ will have $5d^3+d^2\approx 5.37\times 10^9$ nonzeros, which is larger than the number of nonzeros of the original matrix! This is roughly $43$GB of additional memory requirements if $SA$ is stored as a dense double precision array. In addition, a pivoted QR on the large $SA$ is much slower than computing the product $G(SA)$, even if, in theory, both operations require approximately the same number of flops.

\section{Related work\label{sec:related_work}}

The first algorithm that addressed, at least theoretically, the problem of computing the \rsls\ of very large problems was due to Magdon-Ismail and
has complexity $o(nd^2)$ \cite{magdon2010row}.
Drineas et al. in \cite{drineas2012fast} 
improved these results and proposed several fundamental algorithms. In \cite{clarkson2013low_jacm}, Clarkson and Woodruff combine the aforementioned algorithms with novel sparse embeddings to derive the first input-sparsity time leverage scores algorithms. Improvements and lower bounds for such sparse embeddings are studied in
\cite{cohen2016nearly,meng2013low,nelson2013osnap,nelson2013sparsity}. 
Iterative sampling algorithms for leverage scores have been studied in \cite{Cohen:2015cb,Li:2013is}.
Algorithms for general matrices, e.g. when $n\approx d$, are studied in \cite{drineas2012fast,gittens2013revisiting}, approximate the leverage scores of the ``dominant-$k$'' subspace of $A$ using techniques related to power iterations; see \cite{boutsidis2009improved,perry2016augmented} for applications. Bounds for related low-rank approximation from Krylov subspaces are studied in \cite{drineas2018structural}. 
Fast recursive algorithms for the estimation of ridge leverages scores, a regularized version of the leverage scores introduced in \cite{alaoui2015fast}, are studied in \cite{cohen2017input}. Similar ideas are discussed in \cite{kapralov2017single}. 
Parallel aspects of estimating leverage scores via randomized PCA are studied in \cite{Ordozgoiti2017}. Quantum algorithms have also been considered in the literature \cite{Liu:2017ba}.

Another topic related to leverage scores computations is that of estimating the diagonal of a matrix; cf. \cite{bekas2007estimator}. Such estimators have been used in \cite{BekasCurioniFedulova.11} for data uncertainty quantification using iterative methods, mixed precision arithmetic and parallelism. Similar applications and block iterative methods have been studied in \cite{kalantzis2013accelerating}, \cite{Kalantzis2018}. 

Randomized preconditioning for least squares problems has been extensively studied in \cite{avron2010blendenpik,Drineas2010,meng2014lsrn,RokhlinTygert.08}, providing both theoretical guarantees as well as high performance implementations and outperforming state-of-the-art solvers. Similar techniques, including multi-level sketching strategies, have successfully applied to kernel ridge regression \cite{avron2017faster}. Quantifying approximation errors and using this information in applications has also been studied in depth in the context of randomized least squares \cite{lopes2018error}.

Existing randomized leverage scores algorithms have greatly benefited from the fundamental results of Cheung et al on rank estimation and maximal linearly independent column subset selection as a preprocessing step.
Other studies for rank estimation include \cite{saunders2004matrix} as well as the more recent work \cite{ubaru2016fast}. The problem of selecting a subset of linearly independent columns is closely related to the well studied ``Column Subset Selection Problem'' (CSSP), which 
was recently proved to be NP-complete \cite{shitov2021column}. See also \cite{avron2013faster,boutsidis2009improved,civril2014column,deshpande2006adaptive} for state-of-the-art algorithms and hardness results. 

\section{Conclusions\label{sec:conclusions}} 

We have provided algorithms for estimating the statistical leverage scores of rectangular matrices, possibly rank deficient, improving state-of-the-art estimators in terms of complexity and approximation guarantees. 
Our approximation bounds depend on the spectral
gap of the underlying matrix. If the singular values are well separated then we can obtain strong
approximation guarantees for leverage scores that
are useful in practice.  We also developed a set of fast algorithms for rank estimation, column subset selection and least squares preconditioning. Our extensive numerical experiments indicate that our methods perform well in practice on large datasets.

In future work, it would be interesting to study if it is possible to remove
the dependence on the spectral gap and to provide tighter bounds.
Another interesting topic would be to investigate how the proposed methods can be extended for kernel matrices, matrix functions, regularized least squares, and ridge leverage scores computations. 

\section{Acknowledgements\label{sec:acknowledgements}}
We thank the Associate Editor, Michael Berry, and the anonymous reviewers for their valuable feedback. We also thank Petros Drineas for comments on early stages of this work as well as Cristiano Malossi and Christoph Hagleitner for providing access to compute resources. The first author would also like to thank Teodoro Laino for supporting this effort. This work was started in the context of the first author's MS thesis undertaken while at the Computer Engineering \& Informatics Department of the University of Patras, supported in part by an ``Andreas Mentzelopoulos Scholarship''.

\bibliographystyle{siamplain}

\end{document}